\documentclass[pdflatex,sn-mathphys-num]{sn-jnl}% Math and Physical Sciences Numbered Reference Style
\usepackage{multirow}%
\usepackage{amsmath,amssymb,amsfonts}%
\usepackage{mathrsfs}%
\usepackage[title]{appendix}%
\usepackage{xcolor}%
\usepackage{textcomp}%
\usepackage{manyfoot}%
\usepackage{booktabs}%
\usepackage{algorithm}%
\usepackage{algorithmicx}%
\usepackage{algpseudocode}%
\usepackage{listings}%
\usepackage{subcaption}     % To create subtables within a single table environment

\usepackage[utf8]{inputenc} % allow utf-8 input
\usepackage[T1]{fontenc}    % use 8-bit T1 fonts
\usepackage{hyperref}       % hyperlinks
\usepackage{url}            % simple URL typesetting
\usepackage{booktabs}       % professional-quality tables
\usepackage{amsfonts}       % blackboard math symbols
\usepackage{nicefrac}       % compact symbols for 1/2, etc.
\usepackage{microtype}      % microtypography
\usepackage{booktabs}       % For professional-quality table rules
\usepackage{siunitx}        % For aligning numbers by decimal point
\usepackage{graphicx}       % Often useful
\usepackage{amsmath}        % For math symbols
\usepackage{multirow}       % For spanning rows in a table
\usepackage{lineno}

\usepackage{tabularx}
\usepackage{amsfonts}       % blackboard math symbols
\usepackage{lipsum}		% Can be removed after putting your text content
\usepackage{doi}
\usepackage{verbatim}
\usepackage{caption}
\usepackage{array}       % Improves table formatting
\usepackage{colortbl}  
\usepackage[normalem]{ulem}
\usepackage{wrapfig}
\usepackage{diagbox} % for diagonal split cell
\usepackage{float} 
\usepackage{booktabs, tabularx, multirow}
\newcolumntype{Y}{>{\raggedright\arraybackslash}X}
\definecolor{HeaderGray}{gray}{0.90}
\definecolor{OurModelLight}{gray}{0.97}
\definecolor{OurModelDark}{gray}{0.93}
\definecolor{lavender}{RGB}{216,210,245}
\definecolor{lightlavender}{RGB}{240,242,255}
\definecolor{bestblue}{RGB}{54,85,228}
\definecolor{influencer_base}{RGB}{255, 120, 0} % Brighter Orange
\definecolor{influenced_base}{RGB}{0, 100, 200} % Deeper Blue

\definecolor{influencer_light}{RGB}{255, 180, 80}
\definecolor{influenced_light}{RGB}{100, 180, 255}

\definecolor{attention_arrow_color}{RGB}{150, 50, 200} % A distinct purple
\usepackage{tikz}
\usetikzlibrary{arrows.meta,positioning}

\usetikzlibrary{positioning, arrows.meta, shadows.blur, decorations.markings, fadings}
\definecolor{influencer_color}{RGB}{255, 159, 67} % Warm Orange
\definecolor{influenced_color}{RGB}{86, 159, 217} % Cool Blue
\definecolor{arrow_color}{RGB}{88, 88, 88}      % Dark Grey
\theoremstyle{thmstyleone}%
\theoremstyle{thmstyletwo}%

\theoremstyle{thmstylethree}%

\begin{document}

%\title[BrainSymphony]{BrainSymphony: A Lightweight, Modular Transformer-Driven Fusion of fMRI Time Series and Structural Connectivity}
%\title[BrainSymphony]{BrainSymphony: A lightweight modular foundation model integrating structure and function in the human brain}
\title[BrainSymphony]{BrainSymphony: A parameter-efficient multimodal foundation model for brain dynamics with limited data}


% \author*[1,2]{\fnm{First} \sur{Author}}

% \author[2,3]{\fnm{Second} \sur{Author}}

% \author[1,2]{\fnm{Third} \sur{Author}}

\author[1]{\fnm{Moein} \sur{Khajehnejad}}

\author[2]{\fnm{Forough} \sur{Habibollahi}}
\author[1]{\fnm{Devon} \sur{Stoliker}}
\author*[1,5,6]{\fnm{Adeel} \sur{Razi}}\email{adeel.razi@monash.edu}

\affil[1]{\orgdiv{Turner Institute for Brain and Mental Health}, \orgname{School of Psychological Sciences, Monash University}, \orgaddress{\city{Melbourne}, \country{Australia}}}

\affil[2]{\orgdiv{Cortical Labs}, \orgaddress{\city{Melbourne}, \country{Australia}}}

\affil[5]{\orgdiv{Queen Square Institute of Neurology}, \orgname{University College London}, \orgaddress{\city{London}, \country{United Kingdom}}}
\affil[6]{\orgname{CIFAR Azrieli Global Scholars Program}, \orgaddress{\city{Toronto}, \country{Canada}}}


\abstract{Foundation models are transforming neuroscience but are often prohibitively large, data-hungry, and difficult to deploy. Here, we introduce BrainSymphony, a lightweight and parameter-efficient foundation model with plug-and-play integration of fMRI time series and diffusion-derived structural connectivity, allowing unimodal or multimodal training and deployment without architectural changes while requiring substantially less data compared to the state-of-the-art. The model processes fMRI time series through parallel spatial and temporal transformer streams, distilled into compact embeddings by a Perceiver module, while a novel signed graph transformer encodes anatomical connectivity from diffusion MRI. These complementary representations are then combined through an adaptive fusion mechanism. Despite its compact design, BrainSymphony consistently outperforms larger models on benchmarks spanning prediction, classification, and unsupervised network discovery. Highlighting the model's generalizability and interpretability, attention maps reveal drug-induced context-dependent reorganization of cortical hierarchies in an independent psilocybin neuroimaging dataset. BrainSymphony delivers accessible, interpretable, and clinically meaningful results and demonstrates that architecturally informed, multimodal models can surpass much larger counterparts and advance applications of AI in neuroscience.
}




\maketitle

\section{Introduction}\label{intro}
Strong multimodal models that jointly integrate functional and structural neuroimaging are urgently needed yet remain unavailable.
These approaches are central to understanding how dynamic neural activity emerges from, or constrain by, the brain’s structural architecture. Functional magnetic resonance imaging (fMRI) provides temporally resolved signals that reflect ongoing cognitive processes, while structural connectivity, typically derived from diffusion MRI (dMRI), describes the anatomical pathways that constrain and shape these functional dynamics. Together, these modalities offer complementary perspectives: functional signals capture the brain in action, whereas structural networks provide the stable scaffold that supports and limits such activity. Integrating them promises more holistic and biologically grounded models of brain function, with the potential to improve prediction of behaviour and clinical outcomes~\cite{sporns2016networks}. Such integration is particularly important in clinical neuroscience, where robust biomarkers require capturing both transient neural dynamics and the anatomical pathways that constrain them~\cite{suarez2020linking,vieira2017using}. Yet, effectively fusing these distinct data types remains a formidable challenge due to the computational costs and complexity of multimodal modeling. This has limited widespread adoption, leaving most studies reliant on unimodal approaches.

Early attempts relied on handcrafted features, such as correlation-based connectivity matrices~\cite{khajehnejad2023complex}, or shallow models including general linear models and independent component analysis~\cite{beldzik2013contributive}. With the advent of deep learning, graph-based architectures such as BrainNetCNN~\cite{kawahara2017brainnetcnn} and BrainGNN~\cite{li2021braingnn} enabled more expressive modelling of connectome data, later extended by Transformer-based designs like the Brain Network Transformer~\cite{kan2022brain}. While influential, these approaches often reduce the brain to a static graph, overlooking the rich temporal dynamics of neural activity and the nonlinear dependencies that span across spatial and temporal scales~\cite{bullmore2009complex}. Dynamic models such as TAVRNN~\cite{khajehnejad2024tavrnn} have begun to address this by explicitly capturing time-varying interactions, offering a more refined view of neural dynamics.

The latest wave of progress has come from large-scale, self-supervised foundation models for neuroimaging. Inspired by breakthroughs in natural language processing and computer vision~\cite{devlin2019bert,assran2023self}, recent efforts such as BrainLM~\cite{caro2023brainlm} and Brain-JEPA~\cite{dong2024brain} have demonstrated the potential of general-purpose architectures trained directly on fMRI time series. However, these models are often prohibitively large, demanding enormously large datasets and huge computational resources. More critically, they typically focus on either functional or structural data in isolation, leaving the multimodal integration of these intrinsic neural features largely unexplored.

Here, we introduce \textit{BrainSymphony}, a lightweight \emph{modular} foundation model that brings fMRI time series and diffusion-derived structural connectivity into a single, flexible framework. BrainSymphony is explicitly designed to be \emph{plug-and-play}: each modality-specific encoder (spatiotemporal Transformers for fMRI; a signed graph Transformer for structural connectomes) can be used independently or in combination, enabling training and deployment with \emph{fMRI only}, \emph{dMRI-SC only}, or \emph{paired multimodal} data. BrainSymphony is modality-agnostic at deployment: it can operate on whatever data are available.
When both modalities are present, their complementary representations are distilled into compact embeddings and combined through an adaptive fusion mechanism; when one modality is missing, the corresponding pathway can simply be omitted without altering the overall framework.

All fMRI data were preprocessed and parcellated into $n=450$ \textit{regions of interest} (ROIs)—spatially distinct brain areas defined using the Schaefer-400 atlas for cortical regions~\cite{schaefer2018local} and the Tian-Scale~III atlas for subcortical regions~\cite{tian2020topographic}.
This architecture not only achieves state-of-the-art performance across prediction, classification, and unsupervised network discovery, but also provides interpretable insights into brain dynamics. Notably, as a worked example, when applied to a unique external psychedelic neuroimaging dataset~\cite{novelli2025psiconnect}, BrainSymphony revealed drug-induced reorganization of cortical networks through its attention maps that highlight changing dependencies between ROIs over time and indicate which signals the model relies on.

By combining efficiency, interpretability, and neuroscientific validity, BrainSymphony delivers accessible and clinically meaningful results that surpass much larger and more data-hungry counterparts, and provides an architectural basis for future applications of artificial intelligence in neuroscience

\section{Results}\label{results}
BrainSymphony, is a lightweight multimodal foundation model that unifies dynamic brain activity and anatomical structure within a single representational space (Fig.~\ref{fig:schematic}). The framework jointly processes functional MRI (fMRI) time series and diffusion MRI (dMRI)–derived structural connectivity to learn compact, interpretable embeddings at the ROI level. The fMRI stream comprises three complementary encoders: a Spatial Transformer that captures inter-regional dependencies, a Temporal Transformer that captures neural dynamics across time, and a 1D-CNN context extractor that encodes local temporal patterns (Fig.~\ref{fig:schematic}a). Their outputs are distilled by a Perceiver module, which performs cross-attention between high-dimensional fMRI tokens and a set of learned latents to obtain a compact summary of functional activity (Fig.~\ref{fig:schematic}b). In parallel, a Signed Graph Transformer encodes the weighted structural connectome, leveraging edge-aware attention to represent anatomical couplings (Fig.~\ref{fig:schematic}c). Finally, an adaptive gating mechanism fuses functional and structural embeddings, enabling the model to dynamically weigh modalities depending on task demands. For more details, see Section \ref{dataset}. This architecturally informed design provides a data-efficient yet expressive representation of brain organization, serving as the foundation for the analyses presented in the subsequent sections.

\begin{figure}[h!]
    \centering
    \includegraphics[width=0.87\linewidth]{figures/Figure_1.pdf}
    \caption{\textbf{Architecture of BrainSymphony.} \textbf{a)} Three parallel streams encode the fMRI time series: a Spatial Transformer captures relationships between brain regions, a Temporal Transformer models dynamics over time, and a 1D-CNN extracts local features from the signal context.
    \textbf{b)} The outputs from the three fMRI encoder streams are fed into a Perceiver module. This module uses cross-attention against a set of learned latents to distill the rich fMRI information into a compact, fixed-size representation.
    \textbf{c}, In parallel, a Signed Graph Transformer encodes the structural connectome. A learned gate adaptively fuses functional and structural embeddings for downstream prediction.}

    \label{fig:schematic}
\end{figure}

We evaluate BrainSymphony through a series of analyses designed to test both predictive performance and neuroscientific validity. First, we benchmark the model on prediction tasks in the HCP-Aging dataset, comparing unimodal and multimodal variants against established baselines. We then examine parameter efficiency, demonstrating that BrainSymphony achieves state-of-the-art accuracy with orders-of-magnitude fewer parameters than existing foundation models. To assess biological plausibility, we test whether the model’s embeddings recover canonical functional networks in an unsupervised setting. We further validate the fidelity of the learned representations through reconstruction of spatial, temporal, and structural signals. Finally, we apply the model to an independent psilocybin dataset, demonstrating its robust reconstruction capabilities and showing that its attention maps capture drug-induced reorganization of cortical dynamics. Together, these results demonstrate that BrainSymphony is both a high-performing and an interpretable multimodal foundation model for neuroimaging.

\subsection{Multimodal fusion yields consistent gains across benchmarks}

\begin{figure}[H]
    \centering
    \includegraphics[width=1\linewidth]{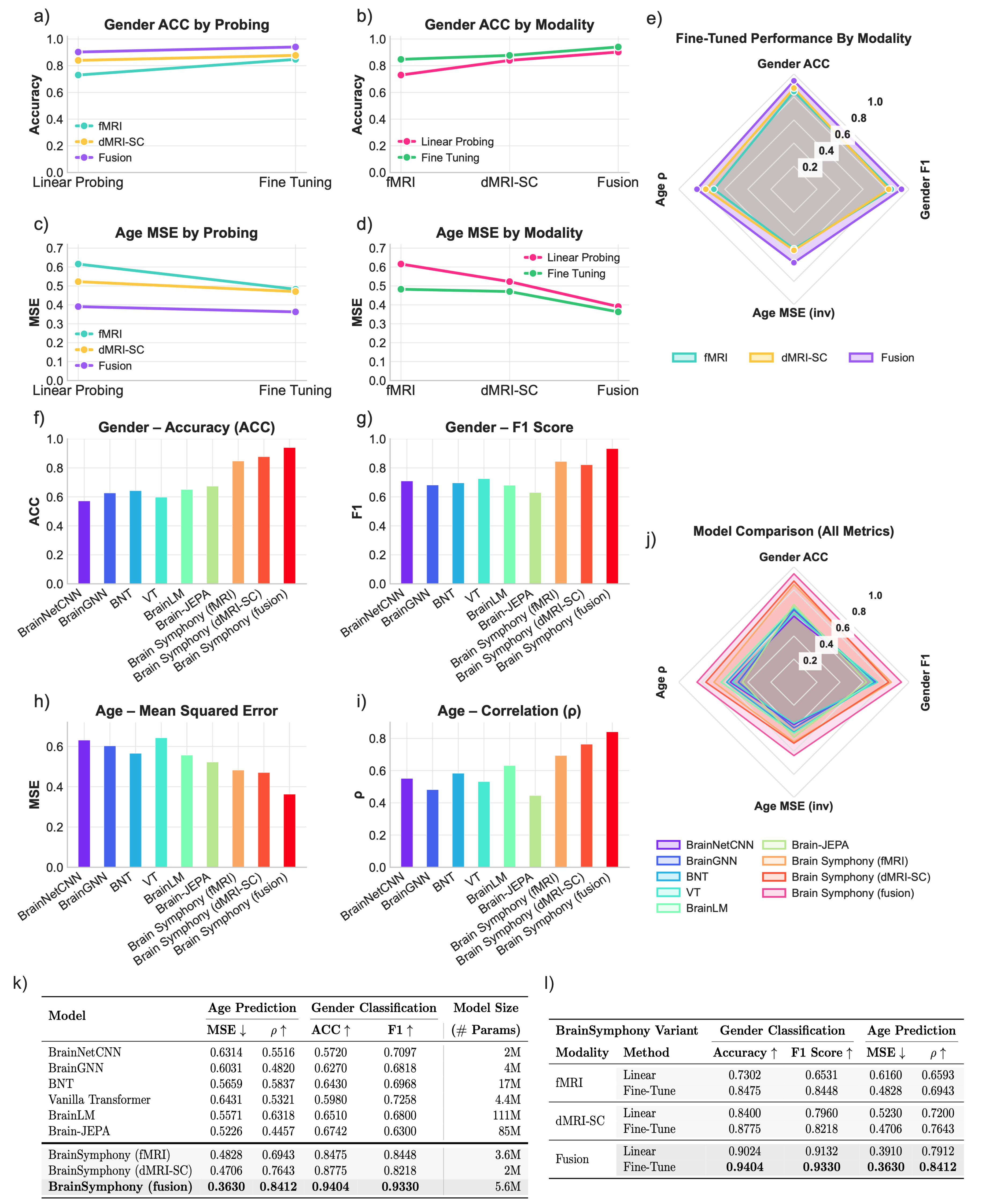}
\caption{\textbf{Performance of BrainSymphony across modalities, tasks, and comparison with baselines.} 
\textbf{a–b)} Gender classification: fine-tuned fusion achieves the highest accuracy (94.0\%) and F1 score ($0.933$). 
\textbf{c–d)} Age prediction: fusion yields the lowest MSE ($0.363$) and highest correlation ($\rho=0.841$). 
\textbf{e)} Radar chart of peak performance (MSE inverted for visualization), showing consistent superiority of the fine-tuned multimodal model across tasks. 
\textbf{f–g)} Gender classification performance of competing models (ACC, F1). 
\textbf{h–i)} Age prediction performance of competing models (MSE, $\rho$). 
\textbf{j)} Overall performance summary: BrainSymphony (fusion) attains the best scores across all metrics while using only $5.6$M parameters, substantially fewer than BrainLM ($111$M) and Brain-JEPA ($85$M). 
\textbf{k)} A detailed breakdown of \textit{BrainSymphony}'s performance across different modalities and evaluation protocols. \textbf{l)} Comparison of our proposed \textit{BrainSymphony} model against baselines, showing key performance metrics alongside model parameter co}

    \label{fig:1}
\end{figure}

We first assessed BrainSymphony on the held-out group of HCP-Aging dataset to evaluate how performance varies by input modality (fMRI, dMRI-based structural connectivity, or their fusion) and by training paradigm (linear probing versus full fine-tuning). The two downstream tasks—gender classification and age prediction—represent complementary challenges: the former is a discrete classification benchmark, while the latter is a continuous regression problem. Together they provide a stringent test of whether multimodal representations improve predictive accuracy across different types of neurocognitive targets. As shown in Fig.~\ref{fig:1}a-e, the fusion of fMRI and structural connectivity embeddings consistently yielded the strongest results, with particularly pronounced improvements when models were fine-tuned end-to-end (detailed report of the performance metrics can be found in the table in Fig.~\ref{fig:1}k.

For gender classification (Fig.~\ref{fig:1}a,b), fine-tuning enhanced performance for all modalities relative to linear probing, underscoring the value of task-specific adaptation even for pretrained representations. Nevertheless, linear probing alone already performs strongly, especially for the fused modality. The multimodal fusion model achieved the highest accuracy ($94.04\%$) and F1-score ($0.933$), substantially surpassing unimodal variants (fMRI-only $84.75\%$ accuracy; dMRI-SC-only $87.75\%$). This demonstrates that structural connectivity provides complementary information to fMRI signals for class-level distinctions, and that the adaptive fusion gate can effectively exploit this synergy. The ability to approach ceiling-level classification performance in such a large and heterogeneous dataset highlights the robustness of the fused representation.

For age prediction (Fig.~\ref{fig:1}c,d), which requires capturing subtle inter-individual differences rather than categorical boundaries, a similar pattern emerged. Fusion model again outperformed the unimodal one, achieving the lowest mean squared error (MSE $=0.363$) and highest correlation with chronological age ($\rho = 0.841$). Notably, the gain from fusion was larger in the regression setting than in classification, suggesting that combining structural and functional signals is especially powerful for modeling continuous phenotypes. These results indicate that BrainSymphony’s fusion variant captures age-related variation more effectively than either fMRI-only or dMRI-SC-only models, consistent with the idea that structural scaffolds constrain functional dynamics across the lifespan.

The summary radar chart and performance table (Fig.~\ref{fig:1}e) provide a holistic comparison across all metrics, making clear that fusion with fine-tuning is consistently superior. Importantly, the architecture’s design explains these improvements: the parallel spatial and temporal transformers extract rich functional embeddings from fMRI, while the Signed Graph Transformer encodes the weighted topology of white-matter connectivity. The adaptive fusion gate then learns to weight these modality-specific contributions differently across regions and individuals, allowing flexible and biologically grounded integration. This stands in contrast to prior multimodal models that rely on static concatenation or simple averaging, which often dilute complementary signals.

Taken together, these results establish two key findings. First, fusion improves generalization across both discrete and continuous behavioural targets, confirming that structural and functional data carry non-redundant information about the brain. Second, fine-tuning is critical: while pretrained representations already carry useful structure, task-specific adaptation unlocks the full potential of multimodal integration. These results demonstrate the novelty and strength of BrainSymphony’s design, providing the first evidence that a lightweight multimodal foundation model can outperform unimodal baselines across a diverse set of prediction tasks in large-scale neuroimaging datasets.

% -----------------------------
% Figure 1
% -----------------------------
% \begin{figure}[!htbp]
%     \centering
%     \includegraphics[width=1\linewidth]{figures/Figure_1.pdf}
%     \caption{\textbf{Fusion with fine-tuning delivers the best performance on HCP-Aging.}
%     Comparison across data modalities (fMRI, dMRI-based structural connectivity, and their fusion) and training paradigms (linear probing vs.\ full fine-tuning).
%     \textbf{a}, Gender classification: fine-tuned fusion achieves the highest Accuracy (94.0\%) and F1 ($0.933$).
%     \textbf{b}, Age prediction: fusion yields the lowest MSE ($0.363$) and highest correlation ($\rho=0.841$).
%     \textbf{c}, Summary table and radar chart of peak performance (MSE inverted for visualization). Across tasks, the fine-tuned multimodal model is consistently superior.}
%     \label{fig:1}
% \end{figure}

\subsection{State-of-the-art performance with orders-of-magnitude fewer parameters}

To contextualize BrainSymphony’s performance, we benchmarked it against a suite of widely used architectures spanning graph neural networks (BrainNetCNN~\cite{kawahara2017brainnetcnn}, BrainGNN~\cite{li2021braingnn}), transformer-based connectivity models (BNT and Vanilla Transformer~\cite{kan2022brain}), and large-scale foundation models (BrainLM~\cite{caro2023brainlm}, Brain-JEPA~\cite{dong2024brain}). Across both classification and prediction tasks in HCP-Aging, the multimodal fusion variant of BrainSymphony decisively outperformed all baselines (Fig.~\ref{fig:1}f-j, table in Fig.~\ref{fig:1}l).

For gender classification (Fig.~\ref{fig:1}f,g), BrainSymphony (fusion) achieved the highest accuracy ($94.04\%$) and F1-score ($0.933$), outperforming BrainLM (Accuracy $=65.10\%$; F1 $=0.680$), Brain-JEPA (Accuracy $=67.42\%$; F1 $=0.630$), and all other baselines. Similarly, for age prediction (Fig.~\ref{fig:1}h,i), BrainSymphony reached the lowest mean squared error and the strongest correlation with chronological age compared to the best-performing baseline BrainLM (MSE $=0.5571$; $\rho=0.6318$). These gains were not marginal: improvements exceeded 20 percentage points in classification accuracy and reduced error by over 35\% in regression. Such across-the-board advances demonstrate that BrainSymphony’s fused multimodal embeddings capture richer and more generalizable brain representations than unimodal or static graph-based models (Fig.~\ref{fig:1}j).

Equally important is the efficiency with which these results are achieved. Whereas BrainLM and Brain-JEPA contain 111M and 85M parameters respectively, BrainSymphony (fusion) uses only 5.6M parameters—over an order of magnitude smaller (table in Fig.~\ref{fig:1}l). This compactness is not a byproduct of under-parameterization but rather a direct consequence of the architecture: spatial and temporal transformers tailored to fMRI dynamics, a Signed Graph Transformer specialized for structural connectivity, and an adaptive fusion mechanism that avoids redundancy by learning where each modality contributes most. In contrast, existing foundation models often scale parameter count indiscriminately, requiring vast datasets and computational resources that limit accessibility and reproducibility.

The combination of superior accuracy and drastic parameter reduction highlights a central novelty of BrainSymphony. It challenges the prevailing assumption that larger foundation models are inherently more powerful, showing instead that neuroscience-inspired architectural priors can unlock both performance and efficiency. This is critical for democratizing foundation models in neuroimaging: a model of BrainSymphony’s size can be trained and fine-tuned on widely available GPU resources, making advanced multimodal representation learning accessible to a broader segment of the community. Furthermore, its lightweight design facilitates deployment in clinical or mobile research settings where computational budgets are constrained.

Together, these results establish BrainSymphony as the first modular, modality-agnostic multimodal foundation model to simultaneously deliver state-of-the-art accuracy and unprecedented parameter efficiency. By showing that carefully designed, domain-aware architectures can outperform massive models by orders of magnitude, BrainSymphony provides a template for the next generation of accessible, interpretable, and high-performing AI in neuroscience.

\subsection{High-fidelity reconstructions of spatial, temporal, and structural signals}

\begin{figure}[H]
    \centering
    \includegraphics[width=1\linewidth]{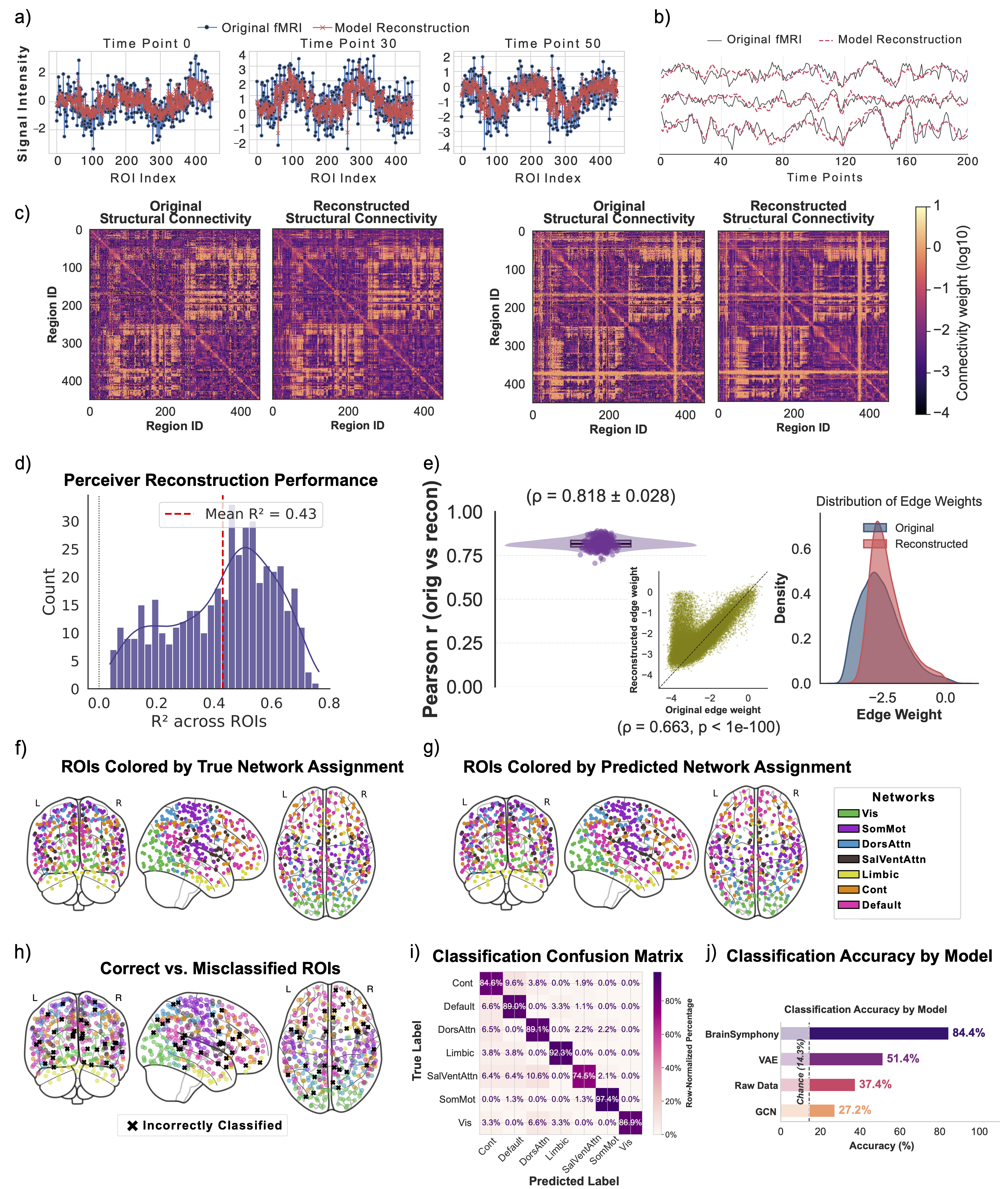}
    \caption{\textbf{Functional network identification and multimodal reconstruction of brain activity and connectivity.}
    \textbf{a)} Reconstructed spatial activation patterns for a sample subject at representative time points from the temporal transformer.
    \textbf{b)} Reconstructed BOLD time courses for sample ROIs, closely tracking the original signals from the spatial transformer.
    \textbf{c)} Original versus reconstructed structural connectivity (with connection weights plotted on a logarithmic scale) for two sample subjects, preserving key modular topology.
    \textbf{d)} Distribution of BOLD time series reconstruction accuracy ($R^2$) from the Perceiver block, showing strong performance across ROIs.
    \textbf{e)} Multi-faceted evaluation of structural connectivity reconstruction, showing high per-subject pattern correlation ($\rho = 0.818 \pm 0.028$, left), accurate prediction of individual edge weights (middle), and an empirically-matched distribution of reconstructed weights (right).
    \textbf{f)} Ground truth assignment of 400 ROIs to seven canonical networks.
    \textbf{g)} Network assignments as predicted by the model's classifier.
    \textbf{h)} Spatial map highlighting misclassified ROIs (black $\times$).
    \textbf{i)} Row-normalized confusion matrix showing high per-network accuracy.
    \textbf{j)} Overall classification accuracy, with BrainSymphony (84.4\%) outperforming comparison models and chance level (14.3\%).
    }
    \label{fig:3}
\end{figure}

An essential property of a foundation model is that its learned embeddings retain sufficient information to faithfully reconstruct the original data. To evaluate this, we assessed BrainSymphony’s ability to reproduce held-out spatial, temporal, and structural signals from its compact latent representations (Fig.~\ref{fig:3}a-e). 

First, the model successfully reconstructed spatial activation patterns across all 400 cortical parcels of a sample subject at multiple sample time points ($t=0, 30, 50$), closely matching the distributed activation structure of the original fMRI volumes (Fig.~\ref{fig:3}a). This demonstrates that BrainSymphony’s temporal transformer stream captures fine-grained patterns of co-activation across distributed brain regions rather than collapsing information into coarse or averaged representations.

Second, BOLD time courses reconstructed for individual ROIs, from the spatial transformer, closely tracked the empirical signals, capturing both slow fluctuations and transient peaks in activity (three sample ROIs presented in Fig.~\ref{fig:3}b). Furthermore, the model's Perceiver block, which fuses the outputs from both the temporal and spatial transformer streams, shows strong efficacy in reconstruction of BOLD time series across the entire HCP-Aging test set, where the model achieved a strong mean reconstruction accuracy ($R^2$) of $0.438 \pm 0.137$ across all ROIs (Fig.~\ref{fig:3}d). As a control, when the time points within each ROI were randomly permuted independently, the reconstruction performance dropped drastically to a mean $R^2$ of $0.030 \pm 0.049$, confirming that the model leverages meaningful temporal dependencies rather than spurious correlations. This quantitative result confirms that the model successfully synthesizes fine-grained spatial patterns and temporal dynamics into a cohesive and accurate representation of whole-brain activity.

Finally, BrainSymphony’s Graph Transformer reconstructed structural connectivity matrices that preserved the modular and block-structured organization observed in the empirical dMRI connectomes (two sample subjects shown in Fig.~\ref{fig:3}c). Key within- and between-network blocks were maintained, underscoring that the graph encoding captures not only overall connectivity strength but also the network-level topology critical for constraining functional dynamics.
A detailed quantitative analysis further validates this reconstruction performance from multiple perspectives (Fig.~\ref{fig:3}e). The model achieved a high subject-wise correlation between original and reconstructed connectomes ($\rho = 0.818 \pm 0.028$), demonstrating its ability to faithfully reproduce each individual's unique connectivity template (left panel). At a more granular level, a significant positive correlation was found between the original and reconstructed weights of individual edges pooled across all subjects ($\rho = 0.663, p < 10^{-10}$), confirming that the model learned a meaningful mapping of connection strengths (middle panel). Notably, the model performed best on stronger edges—those that carry greater network-level influence—while the majority of reconstruction errors arose from weaker, low-weight edges. This pattern suggests that BrainSymphony preferentially captures the most impactful connections governing large-scale communication, while tolerating more variance in minor, less consequential edges. Lastly, the overall probability distribution of the reconstructed edge weights closely matched the empirical distribution, indicating that the model generates connectomes with a high-fidelity, biologically-plausible statistical profile (right panel).

Together, these reconstructions confirm that BrainSymphony embeddings contain high-fidelity information about spatial activation, temporal trajectories, and structural architecture. This property is crucial for both interpretability and downstream utility: it ensures that the model’s latent space is anchored in biologically meaningful features, and that predictive gains do not come at the cost of representational fidelity. In contrast to black-box models where performance can mask loss of structural information, BrainSymphony demonstrates that lightweight multimodal integration can achieve state-of-the-art prediction while preserving the fundamental organization of brain signals across scales.

% % -----------------------------
% % Figure 5 caption 
% % -----------------------------
% \begin{figure}[h!]
%     \centering
%     \includegraphics[width=0.8\linewidth]{figures/Figure_5.pdf}
%     \caption{\textbf{High-fidelity reconstructions of spatial, temporal, and structural signals.}
%     \textbf{a}, Spatial activation patterns across 400 ROIs at sample time points ($t{=}0, 30, 50$).
%     \textbf{b}, BOLD time series reconstructions for two example ROIs (model output closely tracks observed signal).
%     \textbf{c}, Original dMRI-based structural connectivity (left) versus Graph Transformer reconstruction (right), preserving modular and block-structured topology across 450 ROIs (boundaries: dashed lines).}
%     \label{fig:5}
% \end{figure}

% -----------------------------
% Figure 3 
% -----------------------------

\subsection{Recovery of canonical functional networks}

A key test of whether a foundation model has learned meaningful brain representations is its ability to recover known organizational structure without explicit supervision during model training. To test this on the HCP-Aging dataset, we used the 400-region Schaefer functional atlas, which partitions the cortex into seven canonical networks: Control (Cont), Default Mode (Default), Dorsal Attention (DorsAttn), Limbic, Salience/Ventral Attention (SalVenAttn), Somatomotor (SomMot), and Visual (Vis). We then trained a `$k$`-nearest-neighbour (`$k$`-NN) classifier on BrainSymphony’s attention-derived embeddings to assign each of the 400 parcels to one of these networks.

Qualitative comparisons illustrate that predicted network assignments closely matched the ground-truth maps across the cortex (Fig.~\ref{fig:3}f,g), with most regions accurately classified into their canonical networks. Misclassified parcels were sparse and tended to occur at known network boundaries, where functional affiliation is often ambiguous even in empirical studies (Fig.~\ref{fig:3}h). This suggests that the errors may reflect genuine transitional zones rather than model failures. The confusion matrix (Fig.~\ref{fig:3}i) quantitatively confirms this, showing that these few errors were not random but instead concentrated between specific, functionally related networks (e.g., Control and Default Mode, or Dorsal Attention and Salience/Ventral Attention). The latter case is particularly notable, as these networks are tightly coupled in supporting attentional control—one mediating top–down, goal-directed focus and the other detecting salient stimuli and reorienting attention—making their boundaries inherently ambiguous in empirical parcellations \cite{Yeo2011,Power2011}. Moreover, these embeddings dissociated densely interconnected systems (most notably the default-mode) more effectively than correlation-based functional connectivity, which is known to struggle with these systems because of shared global fluctuations, indirect pathways, and spatial adjacency effects \cite{Yeo2011,Power2011,Murphy2017,Smith2011}.

Critically, the model achieved 84.44\% accuracy in this unsupervised classification setting, significantly outperforming alternative approaches based on raw time series (37.44\%), variational autoencoder (VAE) embeddings (51.35\%), and graph convolutional network (GCN) embeddings (27.18\%) (all of which were themselves well above the 14.3\% chance level; Fig.~\ref{fig:3}j). These results demonstrate that the features learned during pretraining intrinsically encode the large-scale network structure of the brain, even though the model was never trained with network labels. Importantly, the substantial performance gap relative to established unsupervised baselines highlights the novelty and strength of BrainSymphony’s integrative architecture in extracting biologically relevant patterns.

Together, these findings establish that BrainSymphony does more than optimize predictive benchmarks: it also internalizes key principles of macroscale brain organization. The ability to recover canonical networks in an unsupervised manner demonstrates that the model’s attention mechanisms capture neurobiologically interpretable structure. This property is critical for both trustworthiness and scientific utility, showing that BrainSymphony can serve not only as a predictive engine but also as a discovery tool for understanding the brain’s intrinsic functional architecture.

% \subsection{Drug-induced reorganization under psilocybin}\label{psiconnect}
\subsection{Validating BrainSymphony readouts on an external cohort: reconstruction and directed attention–based influence}

We next evaluated the ability of BrainSymphony’s fMRI representation learning module to generalize to an unseen cohort by applying it to the external PsiConnect dataset~\cite{stoliker2025psychedelics,novelli2025psiconnect}. PsiConnect comprises fMRI recordings from 62 participants imaged before and after the administration of the mind-altering psychedelic, psilocybin. \textit{Psilocybin} (psilocybin administration; 19 mg oral) and \textit{Baseline} (no psilocybin) scan sessions included eyes-closed rest, meditation, and music conditions, and an eyes-open movie condition. These distinctions allow us to examine both \textit{drug effects} (Psilocybin vs. Baseline) and within-session \textit{condition effects} (Rest, Meditation, Music, and Movie).  

\begin{figure}[H]
    \centering
    \includegraphics[width=0.95\linewidth]{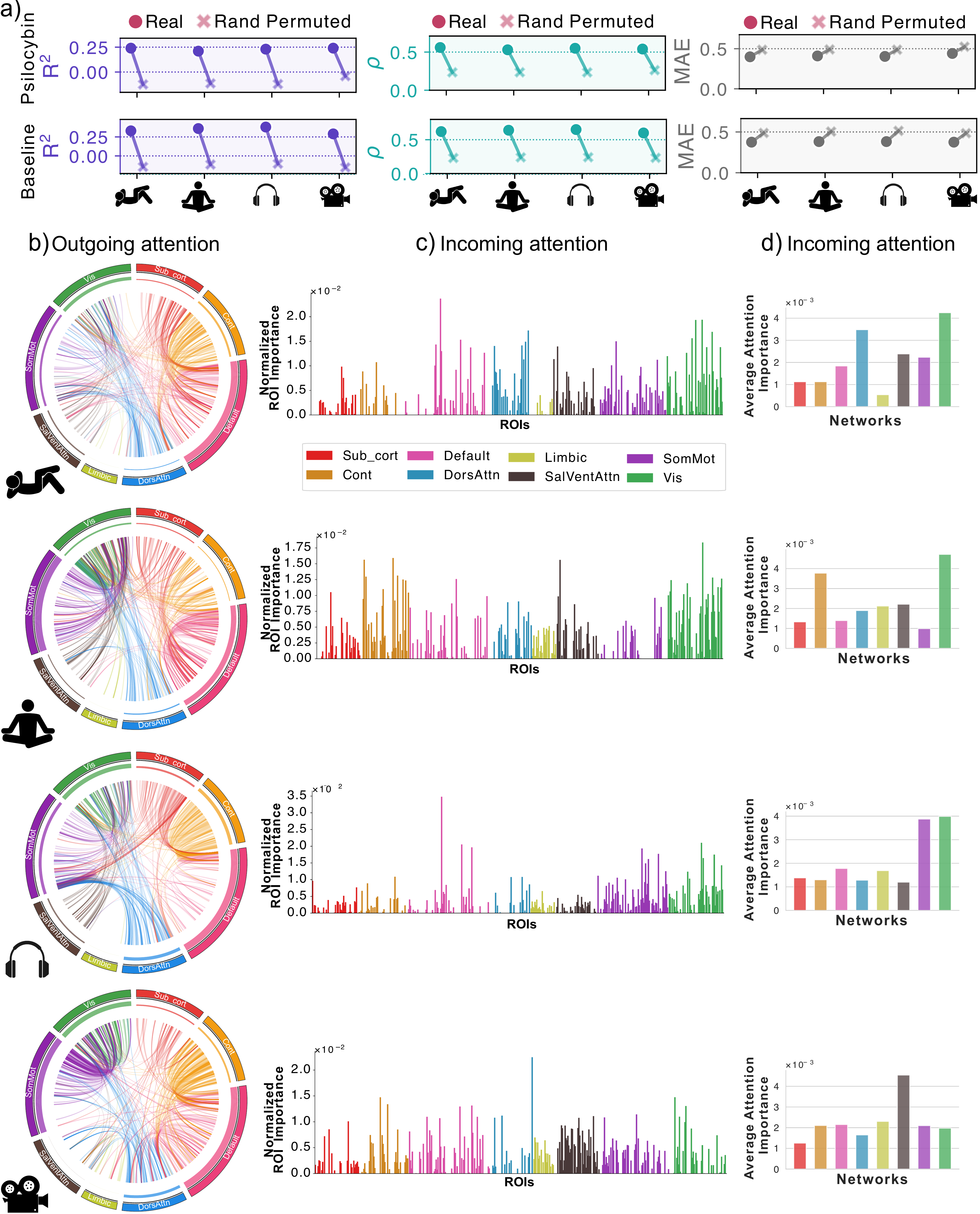}
\caption{\textbf{Reconstruction fidelity and attention mapping of BrainSymphony on the external Psiconnect dataset.} 
\textbf{a)} Reconstruction performance summarized across conditions (Rest, Meditation, Music, Movie) and contexts (Admin vs.\ Baseline). Paired dot plots compare real ROI time series (purple, teal, grey) with randomly permuted controls, showing that BrainSymphony consistently achieves higher $R^2$, $\rho$, and lower MAE. 
\textbf{b)} Circos plots of attention weights (Admin--Baseline difference) for the four conditions, showing the top 500 strongest edges, averaged over all subjects. Each edge is coloured according to the source network, such that all outgoing connections from a given network share the same colour, indicating origin of the attentional query (and thus which network is most strongly integrated with the system). An inner bar track, concentric with the network arcs, depicts the total outgoing attention, with the radial height proportional to this value for each network.
\textbf{c)} Average receptive (incoming) attention to each ROI (Admin–Baseline difference), computed as the mean of all attention weights directed to the ROI, normalized by the total number of ROIs and averaged across subjects. Bars are colored by the ROI’s network affiliation, illustrating preferential drivers of network-wide influence (regions receiving the strongest incoming attention).
\textbf{d)} Network-level averages of incoming attention weights of subplot (c), obtained by pooling across ROIs within each canonical network. 
}
\label{fig:4}
\end{figure}

We first assessed the model’s capacity to reconstruct the fMRI time series from its Perceiver-based fusion embeddings. As summarized in Fig.~\ref{fig:4}a, the model reliably reconstructed ROI time series with substantially higher accuracy than randomly permuted controls (ROI time series permuted independently), as shown by paired comparisons of $R^2$, Pearson $\rho$, and Mean Absolute Error (MAE) across drug vs. no drug effects and imaging conditions. These results demonstrate that BrainSymphony generalizes beyond its training cohort. More detailed results, including full distributions of reconstruction metrics across conditions (Rest, Meditation, Music, and Movie) and drug effects (Psilocybin vs. Baseline), together with representative reconstructed time series, are provided in \nameref{ex_data} Fig.~\ref{fig:S2}.

% This generalization aligns with recent evidence that psilocybin induces widespread reorganization of brain dynamics, reducing modular segregation and increasing global integration~\cite{carhart2016currentbio,10.1101/2021.05.14.444193,10.1038/s41586-024-07624-5}.  

In BrainSymphony, the most comprehensive attention maps are derived from the self-attention layers of the Perceiver block. This module integrates information from the spatial Transformer --capturing dependencies across brain regions--, the temporal Transformer --capturing dynamics across time points--, and the local 1D-CNN context extractor. 
% ROI-specific latent vectors first query the fused spatiotemporal tokens through cross-attention, after which stacked self-attention layers model inter-ROI dependencies within the latent space. 
The resulting self-attention weights provide principled estimates of directed dependencies, distinguishing how strongly each region influences the system (incoming attention) from how strongly each region is influenced by the system (outgoing attention) (see Section~\ref{Perceiver}).

Unlike classical functional connectivity, which typically quantifies pairwise correlations, Perceiver-based attention captures long-range, higher-order relationships across all pairs of regions simultaneously. Crucially, because the receiver of attention is the sender of information---the total incoming attention paid to an ROI---reflects that region’s influence on others (as illustrated in \nameref{ex_data} Fig.~\ref{fig:S3}, Left panel). In contrast, higher outgoing attention indicates how strongly a region is influenced by other regions (Fig.~\ref{fig:S3}, Right panel), as it quantifies the region's active query for information. In this sense, receptive (incoming) attention provides a principled measure of a region's influence on the system, while outgoing attention quantifies the extent to which a region integrates information from the system.

Beyond reconstruction fidelity, we first assessed whether these attention maps encode discriminative information about brain states. Firstly, a linear classifier trained on the full ROI--ROI attention matrices decoded Rest, Meditation, Music, and Movie conditions at baseline (ACC = 0.639, BACC = 0.639, F1 = 0.628), well above chance (25\%). Under psilocybin administration, while still above chance, decoding accuracy declined (ACC = 0.463, BACC = 0.463, F1 = 0.463), consistent with an emergent, drug-dominant signature of desynchronization that reduces categorical structure and generalizes across contexts~\cite{mediano2024effects,siegel2024psilocybin,girn2022serotonergic}. This finding also aligns with prior analyses of the same dataset~\cite{stoliker2025psychedelics}, which reported reduced network modularity and enhanced large-scale network integration under psilocybin. Within the BrainSymphony framework, condition separability in the attention patterns similarly decreased following psilocybin administration, consistent with a shift toward more globally integrated dynamics. Full methodological details and confusion matrices are provided in Supplementary~\ref{attention_decoding} and \nameref{ex_data} Fig.~\ref{fig:S4}.

We next examined how BrainSymphony’s attention mechanisms capture drug- and condition-dependent changes in brain dynamics. As shown in Fig.~\ref{fig:4}b–d, we computed Psilocybin–Baseline differences in the attention maps to characterize how psilocybin alters condition-specific patterns of information flow. The circos plots (Fig.~\ref{fig:4}b) illustrate the average change in outgoing attention across subjects. The inner bar tracks in Fig.~\ref{fig:4}b summarize the total outgoing attention change for each network (excluding within-network connections), with bar height proportional to overall magnitude. These quantify how strongly each network is integrated with (or influenced by) other networks, excluding within-network connections. The central chords show the top 500 strongest outgoing attention edges (within- and between-network). Chord and bar colors denote the source network of each edge, indicating the origin of the attentional query (i.e., which network is most strongly influenced by its targets).

Firstly, inspection of the difference maps (psilocybin minus baseline) reveals that psilocybin administration is associated with a consistent global increase in attention — encompassing region-to-region or bi-directional communication — with no instances of negative values for administration minus baseline, indicating enhanced between- and reduced within-network communication (see \nameref{ex_data} Fig~\ref{fig:S5}). The observed pattern of reduced network integrity in higher-order systems ---particularly within the DMN---together with increased cross-network connectivity and disruption of canonical modular organization, accord with prior reports that psychedelics reduce network segregation and promote reconfiguration of interactions between large-scale networks~\cite{madsen2021psilocybin,petri2014homological,tagliazucchi2016increased}. 
% In contrast, the enhanced within-network communication revealed here (has only been demonstrated with complementary dynamic-network models ~\cite{stoliker2025psychedelics,khajehnejad2024tavrnn}) underscores the novelty of our method, which exposes latent network dynamics that have remained inaccessible to previous analytical approaches. 

% As depicted by these inner bar values (Fig.~\ref{fig:3}b), across all four conditions,
% the Default Mode Network (DMN) consistently shows elevated outgoing attention across imaging conditions—albeit less prominently during meditation—indicating that it remains an important recipient of cross-network influence under psilocybin during internally and externally engaging contexts. The Somato-Motor network also exhibits a marked increase in outgoing attention during meditation and movie, while the Visual network showed the greatest outgoing attention during movie and music. 

As quantified by the inner bar tracks (Fig.~\ref{fig:4}b), the Default Mode Network (pink) consistently shows elevated outgoing attention across all conditions, indicating that it remains a primary recipient of cross-network influence regardless of context. Crucially, the Dorsal Attention network (blue) also exhibits heightened integration, particularly during music and movie. This simultaneous recruitment of typically opposing DMN and DAN systems aligns with well-established findings of reduced network anticorrelation under psilocybin \cite{stoliker2023effective}. Additionally, sensory networks show context-specific surges: the Somatomotor network (purple) exhibits marked increases in outgoing attention during meditation and movie, while the Visual network (green) peaks during music and movie.

Visually, the chords in the Circos plots (Fig.~\ref{fig:4}b) demonstrate stable Control (orange) and Default Mode (pink) sources across contexts, reflecting their sustained role as primary integrators. Crucially, however, the Default Mode network exhibits distinct targeting shifts: directing attention preferentially toward subcortical (red) regions during Rest and Music, focusing inward on self-integration during Meditation, and extending to Dorsal Attention (blue) and subcortical regions during Movie. Music is further distinguished by the emergence of Dorsal Attention and Subcortical sources specifically targeting the Somatomotor network (purple). In contrast, during Meditation and Movie, the Somatomotor network emerges as a prominent integrator. The appearance of thick purple chords originating from this network and extending to visual regions indicates that these conditions shift the Somatomotor cortex from a passive processing role to a state of active cross-modal integration---a known phenomenological feature of psilocybin---by continuously querying the visual system for information.

 %An evaluation of the top 500 strongest chords demonstrates that Control regions exhibit strong increases in outgoing attention under psilocybin during rest and meditation, positioning them as major integrators of input from the rest of the brain. Finally, the Dorsal Attention (DorsAttn) network demonstrates reliably heightened information input in music and movie, consistent with its role in orienting and maintaining attention, pointing to a broader reconfiguration in which both internally directed (DMN) and externally oriented (DorsAttn) systems simultaneously become prominent in whole-brain integration under psilocybin in these conditions. This pattern aligns with prior findings of reduced anticorrelation between these internally and externally oriented networks under psychedelics \cite{stoliker2023effective}.

Next, to assess changes in network influence at the ROI level, we quantified receptive (incoming) attention for each region using the difference between psilocybin and baseline attention maps (Fig.~\ref{fig:4}c). 
Receptive attention was defined as the mean of all attention weights assigned to each ROI by other regions, normalized by the total number of ROIs, and averaged across subjects. Bars are colored according to each ROI’s canonical network affiliation, providing a fine-grained view of which regions become more or less influential in shaping network-wide representations following psilocybin. Receptive attention differences averaged across all subjects for each ROI (Fig.~\ref{fig:4}c) revealed visual ROIs as prominent regions with increased influence during eyes-closed conditions, particularly rest and meditation. 

Pooling across ROIs yielded network-level averages (Fig.~\ref{fig:4}d), which showed systematic differences in the extent to which canonical networks are relied upon, thereby isolating network-specific shifts in information integration under psilocybin.  At the network level condition-dependent increases in received attention were also observed in high-level association networks following psilocybin. Fig.~\ref{fig:4}d shows that the Control network exhibits a significant rise in received attention during meditation, consistent with enhanced top-down regulation and integrative processing associated with meditative states \cite{prakash2025mindfulness}. The Somato-Motor network exhibits a marked increase in received attention specifically during music, possibly reflecting embodied or affective engagement evoked by musical experience \cite{blood2001intensely,koelsch2014brain}, while Salience/Ventral Attention network shows increased attention received during eyes-open Movie. However, the \textit{Visual} network shows the largest increase in attention received from the rest of the brain during all eyes-closed conditions, suggesting psilocybin condition-dependently reorganizes large-scale network communication by increasing the attention directed toward the visual cortex. This pattern indicates that the visual network assumes a more central driving role during eyes-closed states, serving as a primary information source that higher-order systems actively query to reconstruct their own activity. This interpretation accords with studies linking psilocybin to enhanced visual-cortex involvement during experiences of psychedelic-elicited eyes-closed phenomena~\cite{stoliker2025neural,carhart2012neural}.

Together, these converging lines of evidence support the validity of BrainSymphony’s attention maps as biologically interpretable network- and region-level markers of state- and context-dependent changes in brain-wide communication and altered hierarchical balance.

\subsection{Attention-derived directed influence as a mechanistic marker of psilocybin-induced reorganization}\label{psiconnect}
 
\begin{figure}[H]
    \centering
    \includegraphics[width=0.9\linewidth]{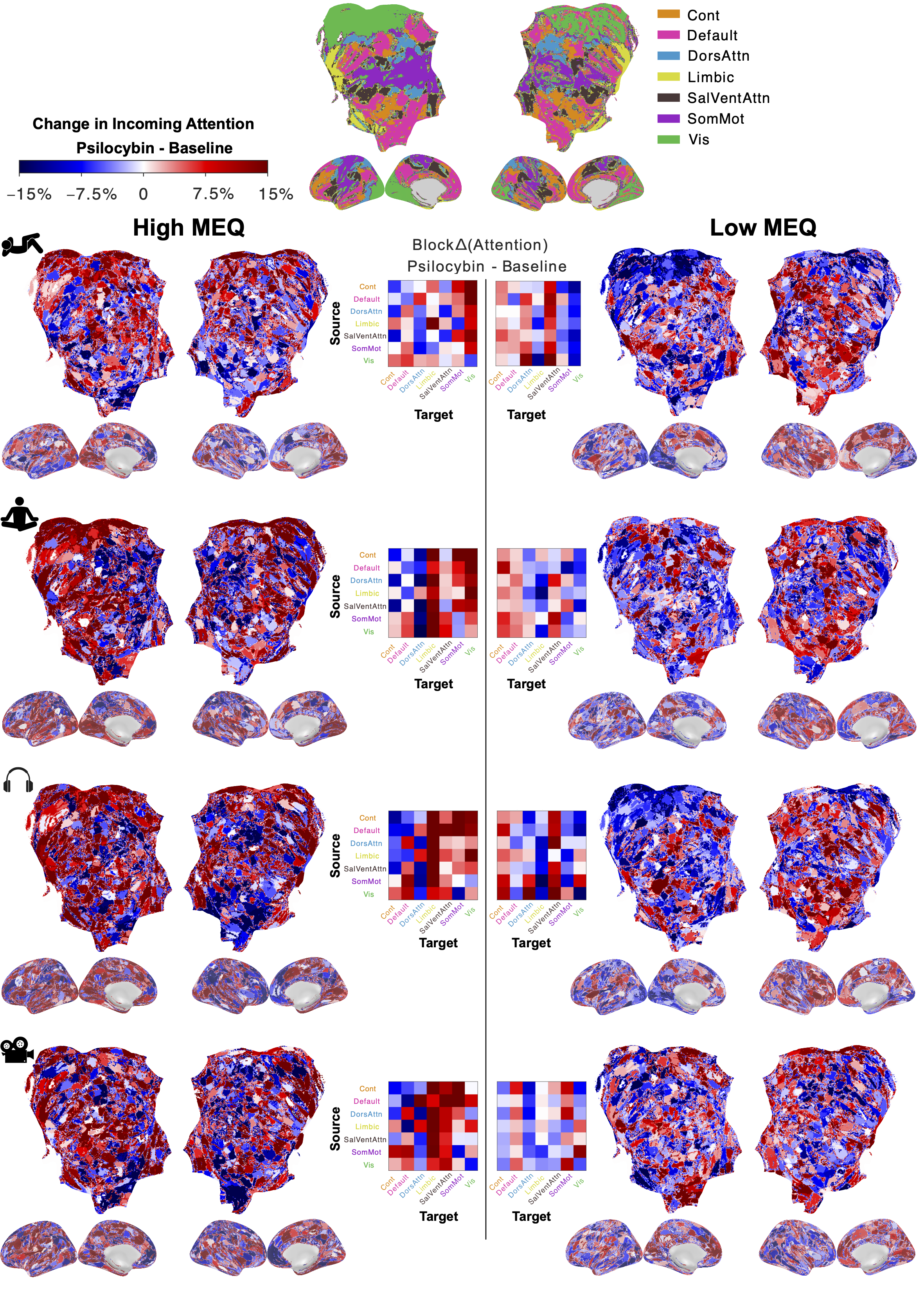}
    \caption{\textbf{Psilocybin-related changes in receptive (incoming) attention across MEQ subgroups and conditions.}
Top row: reference flatmaps showing the seven canonical functional networks. Remaining rows: difference maps (Psilocybin – Baseline) in receptive attention for subjects in the top (High MEQ, left) and bottom (Low MEQ, right) deciles of MEQ30 scores. Each row corresponds to one condition from the PsiConnect dataset (rest, meditation, music, movie). Heatmaps (center) display block-averaged changes between networks, derived from the same subjects, providing a compact summary of psilocybin-related differences in inter-network receptive attention.}
    \label{fig:5}
\end{figure}

We then stratified subjects into high- and low-MEQ groups by selecting the approximate top and bottom 10\% of MEQ30 scores. Specifically, after ranking all subjects by total MEQ30 score, the six highest- and six lowest-scoring individuals were assigned to the high- and low-MEQ groups, respectively. The MEQ30 (Mystical Experience Questionnaire) is a standardized psychometrically validated self-report scale \cite{maclean2012factor,barrett2015validation} administered at the end of the psilocybin session to quantify acute psychedelic effects. It assesses core dimensions of altered experience -- unity, transcendence of time and space, sacredness, positive mood, ineffability and noetic insight -- which load onto a validated four-factor structure (mystical, positive mood, transcendence of time and space, and ineffability). Summed scores across these dimensions index the phenomenological depth of the psychedelic state and are commonly used to predict therapeutic and behavioral outcomes (see \nameref{ex_data} Fig.~\ref{fig:S6} for the distribution of MEQ-30 scores and their mean factor profile \cite{stoliker2025psychedelics}). This stratification allowed us to examine whether subjective intensity of the psychedelic experience modulated network-level reorganization (Fig.~\ref{fig:5}). The flatmaps in this figure depict differences between the Psilocybin and Baseline sessions, thereby localizing regions where receptive (incoming) attention changed under Psilocybin. These different attention maps revealed several robust patterns across conditions and subject groups. First, increases in receptive attention were consistently observed within visual regions in the high-MEQ group. Notably, in the high-MEQ group, this increase was more pronounced during eyes-closed conditions than during the eyes-open movie condition. This pattern aligns with the receptive attention differences observed in the Visual network in (Fig.~\ref{fig:4}d) and suggests that the attention maps track a correspondence between network changes and subjective effect ratings. Notably, we observed less activity during eyes-open viewing, suggesting that the visual systems strong engagement at baseline constrains further drug-induced increases in a Psilocybin minus Baseline contrast. Importantly, these stratified analyses contextualize the group-level patterns seen in Fig.~\ref{fig:4}. Whereas psilocybin generally increased incoming attention to the Visual network in eyes-closed conditions, this reconfiguration was amplified in high-MEQ participants. In these subjects, visual regions acted as dominant sources of brain-wide influence, suggesting that the intensity of the subjective psychedelic experience---characterized by mystical and unitive feelings---scales directly with the degree to which the visual cortex drives global brain dynamics..

Second, the summary heatmaps in Fig.~\ref{fig:5} show that the limbic system showed context- and experience-dependent reorganization. Specifically, meditation, music, and movie all exhibited marked increases in receptive attention to limbic territories under psilocybin compared to baseline, whereas rest did not. This context-specific amplification suggests that limbic circuits—closely tied to emotion, memory, and affective salience—become preferential sources of influence to the rest of the networks dependent on emotionally and perceptually engaging stimuli, indicating the relevance of sensory and psychological context for clinical settings where psilocybin is used. These effects were most pronounced in high-MEQ individuals, suggesting a link between heightened limbic influence on brain networks and the intensity of experiences.

These findings complement the original PsiConnect results, which demonstrated that psilocybin increases integration, reduces modularity, and alters the balance of sensory versus associative systems ~\cite{stoliker2025psychedelics}. Notably, whereas the PsiConnect study analyses showed enhanced separability of imaging conditions in low-dimensional embeddings among subjects with high MEQ scores, BrainSymphony’s attention maps reveal where region- and network-level changes emerge by quantifying shifts in information flow across visual and limbic systems and redistribution of influence across canonical networks.

Taken together, the convergence of these results suggests that psilocybin reorganizes brain dynamics by simultaneously increasing integration and sensitizing the region and network responses to internal and external contexts, here represented by conditions Rest, Meditation, Music, and Movie. BrainSymphony pinpoints the shifting cortical territories across rest, meditation, music, movie attention maps —identifying visual and limbic regions as central sources of influence to the rest of the networks under psilocybin in participants reporting strong subjective experiences (high-MEQ). This attention-based quantification bridges macroscale network reconfiguration with subjective intensity of psychedelic experiences.

For comparison, we performed the same analysis using conventional Pearson correlation–based functional connectivity (see \nameref{ex_data} Fig.~\ref{fig:S7} and Supplementary Section~\ref{fc_comp}). These results showed broad, undifferentiated decreases in connectivity for high-MEQ subjects and increases for low-MEQ subjects, with only limited limbic effects in specific conditions. Unlike BrainSymphony’s attention maps, which capture long-range and higher-order dependencies, functional connectivity provided little mechanistic insight into why subjective intensity differed between groups. This highlights the added value of attention-based receptive inflow measures, which move beyond pairwise correlations to reveal granular network-level redistribution of influence that links specific configurations of brain-wide reorganization more directly to subjective experience. 

Ultimately, these attention maps decode the directionality of the psychedelic state. By distinguishing the sender of queries from the driver of information, BrainSymphony reveals that psilocybin induces a hierarchical inversion: forcing high-order control networks to integrate information broadcasting from a disinhibited visual cortex. This explicitly maps the mathematical logic of the Transformer onto a coherent biological hypothesis of sensory-driven reorganization.

% -----------------------------
% Figure ROI attention 
% -----------------------------

\section{Discussion}\label{discussion}

In this study, we introduced BrainSymphony, a lightweight multimodal foundation model that challenges the emerging view, influenced by large-scale machine-learning practice, that larger architectures are required for superior performance in neuroimaging. Our results show that carefully designed, parameter-efficient models trained on widely available datasets can not only match but also surpass the performance of far larger and more data-intensive approaches. This provides a clear counterpoint to the ``bigger is better'' paradigm that has dominated the development of foundation models in neuroimaging and machine learning more broadly~\cite{caro2023brainlm,dong2024brain}. By embedding architectural priors informed by neuroscience into its design, BrainSymphony demonstrates that efficiency, accuracy, and interpretability can be achieved simultaneously.

Across benchmark tasks in the HCP-Aging dataset, BrainSymphony consistently outperformed graph-based architectures such as BrainNetCNN~\cite{kawahara2017brainnetcnn} and BrainGNN~\cite{li2021braingnn}, transformer-based methods such as the Brain Network Transformer~\cite{kan2022brain}, and recent large-scale models including BrainLM~\cite{caro2023brainlm} and Brain-JEPA~\cite{dong2024brain}. Crucially, these gains were achieved with only 5.6M parameters—over an order of magnitude smaller than BrainLM (111M) and Brain-JEPA (85M). This efficiency arises directly from the architecture: parallel spatial and temporal Transformers tailored to fMRI dynamics, a Signed Graph Transformer specialized for structural connectivity, and an adaptive fusion gate that learns how to combine modalities in a biologically grounded manner. Together, these components enable BrainSymphony to capture multimodal dependencies more effectively than indiscriminate parameter scaling.

The neuroscientific validity of BrainSymphony’s embeddings is underscored by their ability to recover canonical functional networks in an unsupervised setting~\cite{sporns2016networks}. Attention-based embeddings achieved 84.44\% accuracy in network classification, substantially surpassing raw time-series features, graph convolutional networks, and variational autoencoders. This indicates that BrainSymphony encodes large-scale principles of functional brain organization. Importantly, attention maps do not simply replicate functional connectivity: whereas classical connectivity measures capture pairwise correlations, Transformer attention models long-range, higher-order dependencies across all pairs of regions simultaneously, spanning both temporal and spatial dimensions. This property moves BrainSymphony beyond black-box prediction, offering interpretable mechanisms that generate neuroscientifically meaningful hypotheses.

Application to the external PsiConnect dataset~\cite{stoliker2025psychedelics,novelli2025psiconnect} provided an additional test of generalizability and interpretability. BrainSymphony’s receptive attention maps revealed a redistribution of influence such as increased influence of visual regions (Fig.~\ref{fig:3}c–d), and heightened limbic influence on other networks in participants reporting stronger subjective effects—particularly during more engaging contexts such as music listening and movie watching (Fig.~\ref{fig:4}). Notably, the original analysis of the PsiConnect dataset showed that psilocybin makes context-dependent brain dynamics more separable in low-dimensional embeddings \cite{stoliker2025psychedelics}. BrainSymphony complements this by providing a mechanistic explanation: separability arises from reweighting of information outflow across visual and limbic systems, with these systems emerging as dominant sources of long-range influence particularly in participants reporting stronger subjective effects. Together, these perspectives suggest that psilocybin reorganizes large-scale brain dynamics by enhancing integration, suggesting a mechanistic bridge between macroscale network changes and subjective phenomena such as internally generated perception and heightened affect.

The broader implications of these findings are threefold. First, BrainSymphony demonstrates that it is possible to democratize access to high-performing neuroimaging foundation models. By delivering state-of-the-art results with an order of magnitude fewer parameters, it reduces the computational barrier for research groups without access to massive GPU clusters, thereby making advanced multimodal analysis more widely accessible. Second, its architectural modularity enables flexible deployment: components such as the Signed Graph Transformer can be adapted to matrices of effective connectivity, including those derived from Dynamic Causal Modeling (DCM)~\cite{friston2003dynamic}. Third, the interpretability of BrainSymphony’s attention mechanisms provides a powerful avenue for hypothesis generation in both systems neuroscience and psychopharmacology, linking representational changes to brain state, subjective experience, and potential clinical outcomes.

For future users aiming to apply the pretrained BrainSymphony model to new datasets, it is essential to follow the same preprocessing pipeline described in Section~\ref{dataset}. Specifically, input fMRI time series should be robustly scaled (median-centered and divided by the interquartile range) across participants for each ROI, and data must use the same parcellation scheme—comprising 50 Tian subcortical regions followed by 400 Schaefer cortical regions (ROI order: 1–50 = Tian-Scale III subcortex; 51–250 = Schaefer-400 left hemisphere; 251–450 = Schaefer-400 right hemisphere). Structural connectivity matrices should be constructed as streamline counts normalized by ROI volume, subsequently log\textsubscript{10}-transformed and aggregated into the 450-ROI parcellation.
Depending on the downstream task (e.g., regression, classification, or reconstruction), the pretrained model can then be fine-tuned via the appropriate output head while keeping the core multimodal encoders fixed. Notably, BrainSymphony is modular: either the fMRI (functional) or structural component—or both jointly—can be used, depending on data availability and the application context (see Supplementary Section \ref{sup_dataset}).

Some limitations warrant consideration. The model was primarily validated on healthy adult cohorts, and its generalizability to developmental or clinical populations remains to be established. While attention maps provide interpretable insights into model mechanisms, they are not direct measures of neural causality and should be interpreted with caution.

Looking forward, applying BrainSymphony to diverse clinical datasets is a particularly promising avenue, with potential to aid diagnosis, track disease progression, and predict treatment response in psychiatric and neurological disorders~\cite{vieira2017using}. Extending the Signed Graph Transformer to effective connectivity would allow hypothesis-driven investigations of directional interactions between brain regions, bridging data-driven embeddings with mechanistic models of brain function. Finally, the combination of multimodal integration, architectural efficiency, and mechanistic interpretability positions BrainSymphony as a template for future foundation models in neuroscience: models that are not only powerful predictors, but also transparent tools for discovery. By aligning performance with interpretability, BrainSymphony provides a pathway towards clinically meaningful, computationally accessible, and neuroscientifically grounded AI for the human brain.

\section{Methods}\label{methods}

\subsection*{Overview}
BrainSymphony is a lightweight multimodal foundation model that integrates dynamic brain activity (fMRI) and anatomical architecture (dMRI-derived structural connectivity) into unified ROI-level embeddings (Fig.~\ref{fig:schematic}). The model comprises: (i) a \emph{Spatio–Temporal fMRI encoder} with dedicated Spatial and Temporal Transformers plus a local Context extractor; (ii) a \emph{Perceiver-based fusion} that distills high-dimensional spatiotemporal tokens into compact ROI-wise latents; (iii) a \emph{Signed Graph Transformer} that encodes the weighted structural connectome (while the architecture is also capable of handling signed and directed graphs); and (iv) an \emph{adaptive gating} module that fuses functional and structural embeddings for downstream tasks.

\vspace{4pt}
In the following, we first describe the utilized datasets and preprocessing steps, then the fMRI encoders and Perceiver fusion, followed by the structural encoder and multimodal fusion. We conclude with objectives, training protocols, and baseline methods. Architectural details and hyperparameters are summarized in Table \ref{tab:arch_summary} in the Supplementary Section \ref{hyperparams}.

\subsection*{Datasets and preprocessing}\label{dataset}

For self-supervised pretraining, we utilized data from two large-scale, public datasets: the \textbf{Human Connectome Project Young Adult (HCP-YA)}~\cite{elam2021human} and \textbf{HCP-Aging}~\cite{elam2021human}. A unique psilocybin neuroimaging dataset (\textbf{PsiConnect}) \cite{stoliker2025psychedelics,novelli2025psiconnect} was also used as unseen external data in the test phase. For pretraining, we used resting-state fMRI (rs-fMRI) data from the HCP-YA dataset (967 healthy participants, aged 22-35) and 40\% of the HCP-Aging dataset (262 healthy participants, aged 36--100\texttt{+}). The remaining 60\% of the HCP-Aging dataset, consisting of 394 participants, was held-out as a test set for downstream evaluation tasks. During the fine-tuning and linear probing stages, this held-out data was partitioned into a 6:2:2 ratio for training, validation, and testing, respectively. 
Additionally, we utilized the PsiConnect dataset as an external test set, comprising fMRI data from 62 individuals scanned both before and after psilocybin administration. Quality-control procedures (e.g., excessive head motion and non-uniform scan lengths across participants) resulted in the exclusion of eight datasets, leaving 54 participants for analysis. This unique dataset provided a valuable opportunity to investigate drug-induced alterations in brain dynamics.

Both the HCP-YA and HCP-Aging datasets feature multi-band acquisition with a high temporal resolution of approximately 0.72 seconds. PsiConnect dataset, which was utilized for external validation, features a temporal resolution (TR) of 0.91 seconds. On this external cohort, attention maps derived from our model were studied in an unsupervised manner to elucidate changes in brain dynamics as a result of psilocybin administration.

All fMRI data was preprocessed and parcellated into $n=450$ regions of interest (ROIs) using Schaefer-400 \cite{schaefer2018local} for cortical regions and Tian-Scale III \cite{tian2020topographic} for subcortical regions. For the task of recovering canonical functional networks, following the Schaefer functional atlas, the brain network is partitioned into seven major functional subnetworks: control (Cont), default mode (Default), dorsal attention (DorsAttn), limbic (Limbic), salience/ventral attention (SalVenAttn), somatomotor (SomMot), and visual (Vis). Additionally, we include the 50 subcortical regions defined by the Tian-Scale III parcellation, which we collectively refer to as Sub\_Cort. 

Similar to previous work, we applied robust scaling by subtracting the median and dividing by the interquartile range (IQR) of each ROI’s time series, computed across all participants \cite{caro2023brainlm}. The model input consisted of 450 ROIs, each represented by 200 time points, forming a tensor of shape $R=450 \times T=200$ (see Supplementary Section~\ref{sup_fMRI}).

Diffusion MRI data were preprocessed using MRtrix3, including denoising, motion and eddy-current correction, bias-field correction, and spherical deconvolution tractography. Structural connectivity matrices were derived as log\textsubscript{10}-transformed streamline counts normalized by ROI volume, aggregated into the 450-ROI parcellation (see Supplementary Section \ref{sup_SC}). 

\subsection{Spatio-Temporal Transformer Framework for fMRI Representation Learning}
\label{sec:spatiotemporal}

Functional MRI signals present a formidable challenge for representation learning: they are high-dimensional, noisy, and exhibit intricate dependencies across both space and time. To address these complexities, BrainSymphony factorizes the encoding process into three complementary modules (Fig.~\ref{fig:schematic}a): a \emph{Spatial Transformer} that learns relationships among regions of interest (ROIs), a \emph{Temporal Transformer} that models global temporal dependencies, and a \emph{local Context Extractor} based on 1D convolutional filters that captures fine-grained signal dynamics. Each module is optimized through masked reconstruction, encouraging the model to infer missing information from distributed contextual cues.

The Spatial Transformer treats each ROI as a token and learns its anatomical identity and interactions with other regions by reconstructing masked ROI signals. This design captures the principle that brain regions do not operate in isolation but are embedded in a web of structural and functional dependencies. In parallel, the Temporal Transformer treats each time point as a token and learns long-range dependencies across the full time course, guided by sinusoidal positional encodings that preserve temporal order. To complement these global mechanisms, the Context Extractor applies localized convolutional filters to each ROI’s time series, enabling the model to recognize transient bursts, oscillations, and other fine-scale dynamics often overlooked by global attention mechanisms.

Integrating the outputs of these three encoders poses a major computational challenge: the full spatiotemporal signal contains $R\times T$ tokens, far too large for direct attention. To overcome this, BrainSymphony employs a Perceiver-style latent bottleneck (Fig.~\ref{fig:schematic}b). A fixed set of $R$ learnable latent vectors—one per ROI—cross-attend to the fused spatiotemporal token space, selectively distilling the most informative features into compact ROI-level embeddings. These latents not only reconstruct the original signals with high fidelity but also serve as \emph{neural signatures} for each ROI, capturing spatial context, temporal dependencies, and local dynamics in a unified representation. The resulting ROI embeddings form the functional branch of BrainSymphony and constitute the foundation for downstream decoding and multimodal fusion.

\subsubsection{Formulation of fMRI representation learning}

We formalize the task of functional representation learning as follows. Let $\mathbf{X} \in \mathbb{R}^{B \times R \times T}$ denote a batch of preprocessed fMRI recordings, where $B$ is the batch size, $R$ the number of regions of interest (ROIs), and $T$ the number of sampled time points. Each sample $\mathbf{X}_b \in \mathbb{R}^{R \times T}$ thus represents the spatiotemporal activation profile of a single subject.

The goal of the fMRI encoder is to map $\mathbf{X}$ into compact ROI-level embeddings 
\[
\mathbf{E}_{\mathrm{fMRI}} \in \mathbb{R}^{B \times R \times d_f},
\]
where $d_f$ is the latent dimensionality. These embeddings are trained under a masked reconstruction regime: subsets of ROIs or time points are withheld, and the model is optimized to infer the missing signals from observed context. This objective encourages distributed representations that capture both intra-regional temporal dynamics and inter-regional spatial dependencies. 

The resulting ROI-level embeddings act as \emph{subject-specific neural signatures}, rich enough to support downstream tasks such as classification, regression, and functional network recovery, and ultimately serve as the functional input to the multimodal fusion module of BrainSymphony.

Building on this formulation, we next describe the Spatial Transformer encoder (Fig.~\ref{fig:schematic}a, top), which embeds each ROI as a token and learns context-aware spatial representations guided by gradient-informed positional priors.

% \subsubsection*{Spatial Transformer with gradient-informed positions}
% Each ROI $r$ is a token. We project its time series $\mathbf{x}_r\!\in\!\mathbb{R}^{T}$ to an embedding $\mathbf{z}_r=\mathbf{W}_{\mathrm{embed}}\mathbf{x}_r\in\mathbb{R}^{D}$. To encourage contextual inference, we mask a proportion $p$ of ROI tokens and reconstruct them from the unmasked set (masked-modeling objective following BERT/MAE~\cite{devlin2019bert,he2022masked}).

% \textit{Gradient-informed positional encoding.} To inject neurobiologically meaningful spatial priors, we compute functional \emph{gradients} by diffusion map embedding of ROI connectivity profiles (see Supplementary~\S S2 for full derivation) and align subject-wise gradients to a common template via Procrustes. For ROI $r$, we concatenate its gradient coordinates with a learned positional vector and project to $D$ to obtain the position $\mathbf{P}_r$. Tokens are $\mathbf{z}'_r=\tilde{\mathbf{z}}_r+\mathbf{P}_r$. This places attention over a continuous, functionally relevant manifold rather than arbitrary indices~\cite{margulies2016situating,vos2020brainspace}.

% A stack of $L$ Transformer blocks (MSA+MLP with pre-LN) yields hidden states $\{\mathbf{h}_r\}$, decoded to $\hat{\mathbf{x}}_r=\psi(\mathbf{h}_r)$ with an MLP. The loss is masked MSE over masked ROIs:
% \[
% \mathcal{L}_{\mathrm{spatial}}=\frac{1}{\sum_r (1-M_r)}\sum_r (1-M_r)\|\hat{\mathbf{x}}_r-\mathbf{x}_r\|^2.
% \]

\subsubsection{Spatial Transformer Encoder with Gradient-Informed Positional Encoding}

To capture spatial dependencies in fMRI data, BrainSymphony employs a Spatial Transformer that treats each region of interest (ROI) as a token and learns its interaction-aware representation through self-attention across all regions (Fig.~\ref{fig:schematic}a, top). The encoder is trained under a masked modeling regime: subsets of ROIs are hidden and reconstructed from unmasked ones, forcing the model to infer missing dynamics from distributed context. This strategy, inspired by BERT~\cite{devlin2019bert} and masked autoencoders~\cite{he2022masked}, is particularly suitable for neuroimaging, as it mimics partial observability of the brain and encourages the model to internalize inter-regional dependencies.

\textbf{Tokenization and masking.}  
Given an fMRI sample $\mathbf{X} \in \mathbb{R}^{R \times T}$, each ROI time series $\mathbf{x}_r \in \mathbb{R}^T$ is mapped into the Transformer’s latent space by a learnable projection:
\[
\mathbf{z}_r = \mathbf{W}_{\mathrm{embed}} \mathbf{x}_r, \qquad \mathbf{W}_{\mathrm{embed}} \in \mathbb{R}^{d_f \times T}.
\]
Here, $\mathbf{W}_{\mathrm{embed}}$ is a trainable linear layer analogous to word embeddings in NLP, ensuring that each ROI token is represented in a shared $d_f$-dimensional space suitable for attention. To enforce contextual learning, a binary mask $\mathbf{M} \in \{0,1\}^R$ selects a proportion $p$ of ROIs to be hidden, and their embeddings are replaced by a shared mask token $\mathbf{z}_{\mathrm{mask}} \in \mathbb{R}^{d_f}$:
\[
\tilde{\mathbf{z}}_r = M_r \cdot \mathbf{z}_r + (1 - M_r) \cdot \mathbf{z}_{\mathrm{mask}}.
\]
The network must then reconstruct masked ROI signals from the remaining context, teaching it to capture redundancy and long-range correlations in brain activity.

\textbf{Gradient-Informed Positional Encoding.}
To embed biologically grounded spatial structure into ROI tokens, we introduce a hybrid positional encoding strategy that fuses two complementary components: (1) data-driven coordinates from functional connectivity gradients, and (2) shared learnable embeddings. This hybrid design allows the model to incorporate both subject-specific anatomical variability and population-level spatial priors.

The gradient-based component derives from the intrinsic geometry of functional brain networks. Specifically, we compute a low-dimensional embedding of the brain’s functional organization using \textit{diffusion map embedding}~\cite{coifman2006diffusion}, applied to a symmetric affinity matrix $\mathbf{A} \in \mathbb{R}^{R \times R}$ constructed from pairwise cosine distances between ROI-wise functional connectivity vectors. Each ROI $i$ is described by its connectivity profile $\mathbf{c}_i \in \mathbb{R}^R$, and the affinity between ROIs $i$ and $j$ is defined as:
\begin{equation}
\mathbf{A}_{ij} = 1 - \frac{1}{\pi} \cos^{-1} \left( \frac{\mathbf{c}_i^\top \mathbf{c}_j}{\|\mathbf{c}_i\| \cdot \|\mathbf{c}_j\|} \right),
\end{equation}
which ensures symmetry and bounded similarity in $[0,1]$ based on functional similarity.

To construct the diffusion operator, we first compute the degree matrix $\mathbf{D} \in \mathbb{R}^{R \times R}$ with entries $\mathbf{D}_{ii} = \sum_j \mathbf{A}_{ij}$ and then form the \textit{symmetric normalized graph Laplacian}:
\begin{equation}
\mathbf{L}_{\text{sym}} = \mathbf{I} - \mathbf{D}^{-1/2} \mathbf{A} \mathbf{D}^{-1/2}.
\end{equation}
The eigendecomposition of $\mathbf{L}_{\text{sym}}$ yields:
\begin{equation}
\mathbf{L}_{\text{sym}} \mathbf{U} = \mathbf{U} \boldsymbol{\Lambda},
\end{equation}
where $\mathbf{U} \in \mathbb{R}^{R \times R}$ contains the eigenvectors and $\boldsymbol{\Lambda}$ is a diagonal matrix of eigenvalues. We construct the gradient matrix $\mathbf{G} \in \mathbb{R}^{R \times K}$ by selecting the $K$ nontrivial eigenvectors corresponding to the smallest non-zero eigenvalues (i.e., excluding the constant first eigenvector). The resulting $\mathbf{G}$ provides a smooth, low-dimensional embedding that captures the intrinsic functional geometry of the brain, organizing ROIs along continuous gradients of connectivity.

Each row $\mathbf{g}_i \in \mathbb{R}^K$ of $\mathbf{G}$ encodes the intrinsic functional position of ROI $i$ in the brain's latent manifold, revealing axes such as unimodal-to-transmodal organization. 

\begin{figure}[h!]
    \centering
    \includegraphics[width=1\linewidth]{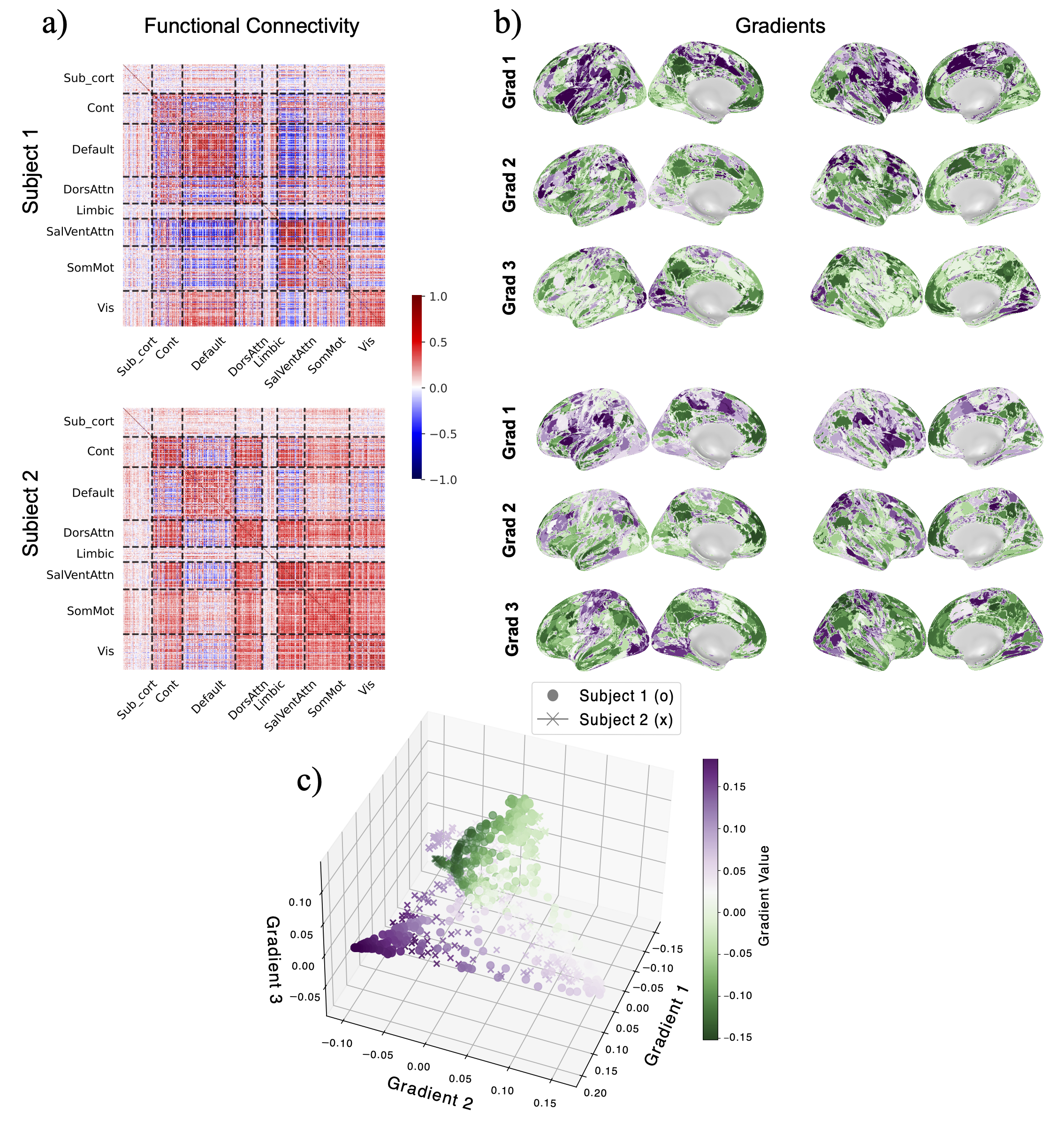}
    \caption{\textbf{Gradient-Informed Positional Encoding.}
\textbf{a)} Subject-specific functional connectivity matrices for two example individuals, computed using Pearson correlation across ROIs. ROIs are grouped by canonical networks from Schaefer-400 and Tian-Scale III atlases (see~\ref{dataset}). 
\textbf{b)} The first three subject-specific functional gradients derived from diffusion embedding of the affinity matrix constructed from ROI-level functional connectivity profiles. These gradients reflect the intrinsic geometry of brain organization, capturing axes such as unimodal-to-transmodal hierarchy.
\textbf{c)} 3D embedding of ROI coordinates for two subjects in the shared gradient space after Procrustes alignment.
}

    \label{fig:Gradient_position}
\end{figure}

However, since functional gradients are computed separately for each subject, the resulting coordinate spaces may be arbitrarily rotated or reflected across individuals due to the inherent sign and rotation ambiguity of eigendecomposition. This misalignment poses a significant challenge when gradients are used as positional encodings in a shared model, as corresponding ROIs across subjects may lie in incongruent embedding spaces. To ensure anatomical and functional correspondence across subjects, we perform gradient alignment using a Procrustes-based approach. Specifically, we iteratively align all subject-level gradients into a common latent space by minimizing inter-subject variance through repeated orthogonal transformations. This procedure enables consistent and biologically meaningful positional encoding across the population, facilitating more effective learning and generalization in downstream tasks (Figure~\ref{fig:Gradient_position}). 

Although these gradients capture rich individual-level topographies, they may underrepresent invariant or abstract spatial priors. To address this, we concatenate each ROI’s gradient-derived embedding $\mathbf{g}_r$ with a shared learnable positional token $\mathbf{p}_r^{\text{learn}}$, and project the result into the Transformer’s latent space using a fusion network:
\[
\mathbf{P}_r = \text{MLP}\left( \left[ \mathbf{g}_r \mathbf{W}_g \,\|\, \mathbf{p}_r^{\text{learn}} \right] \right), \quad \text{s.t.} \quad \mathbf{P}_r \in \mathbb{R}^{d_f},  \mathbf{W}_g \in \mathbb{R}^{K \times d_f}.
\]
This fused representation combines individualized anatomical-functional positioning with task-general positional priors, enabling the model to learn representations that are both neurobiologically meaningful and generalizable across subjects.

Importantly, gradient-informed positional encoding aligns with neuroscientific evidence showing that functional gradients reflect the brain's hierarchical organization across sensory, cognitive, and associative systems~\cite{margulies2016situating, vos2020brainspace}. By leveraging this principle, our model embeds ROIs into a topological space where spatial attention mechanisms operate over continuous, functionally-relevant coordinates rather than arbitrary indices.

% \textbf{Transformer architecture and reconstruction.}  
% Position-enriched embeddings $\mathbf{z}'_r = \tilde{\mathbf{z}}_r + \mathbf{P}_r$ are passed through a stack of $L$ Transformer blocks with multi-head self-attention and feedforward layers, producing contextualized ROI states $\{\mathbf{h}_r\}$. Each state is decoded by an MLP $\psi: \mathbb{R}^D \rightarrow \mathbb{R}^T$ to reconstruct the ROI’s time series:
% \[
% \hat{\mathbf{x}}_r = \psi(\mathbf{h}_r).
% \]
% The masked reconstruction loss is
% \[
% \mathcal{L}_{\mathrm{spatial}} = \frac{1}{\sum_r (1 - M_r)} \sum_{r=1}^{R} (1 - M_r)\,\|\hat{\mathbf{x}}_r - \mathbf{x}_r\|^2,
% \]
% which encourages the model to learn spatial embeddings that accurately predict hidden regional dynamics from distributed information.

\textbf{Transformer architecture.}
The masked ROI embeddings $\tilde{\mathbf{z}}_r$ are enriched with their corresponding positional encodings:
\[
\mathbf{z}_r' = \tilde{\mathbf{z}}_r + \mathbf{P}_r.
\]
The resulting token sequence $\{\mathbf{z}_r'\}_{r=1}^R$ is passed through a stack of $L$ Transformer layers, each consisting of multi-head self-attention and position-wise feedforward networks with residual connections and layer normalization:
\begin{align}
\mathbf{h}_r^{(l+1)} &= \mathbf{h}_r^{(l)} + \text{Dropout}\left(\text{MSA}\left(\text{LN}(\mathbf{h}_r^{(l)})\right)\right), \\
\mathbf{h}_r^{(l+1)} &= \mathbf{h}_r^{(l+1)} + \text{Dropout}\left(\text{MLP}\left(\text{LN}(\mathbf{h}_r^{(l+1)})\right)\right),
\end{align}
where $\mathbf{h}_r^{(0)} = \mathbf{z}_r'$.

\textbf{Reconstruction objective.}
The final hidden states $\{\mathbf{h}_r\}_{r=1}^R$ are decoded through an MLP $\psi: \mathbb{R}^{d_f} \rightarrow \mathbb{R}^T$ to reconstruct the ROI time series:
\[
\hat{\mathbf{x}}_r = \psi(\mathbf{h}_r).
\]
Training is performed using a masked mean squared error (MSE) loss, computed only over masked ROIs:
\[
\mathcal{L}_{\text{spatial}} = \frac{1}{\sum_r (1 - M_r)} \sum_{r=1}^{R} (1 - M_r) \cdot \|\hat{\mathbf{x}}_r - \mathbf{x}_r\|^2.
\]
This objective encourages the model to infer missing regional dynamics using information from spatially and functionally related regions, fostering representations that reflect both local circuit-level activity and global cortical topology.

By treating ROIs as tokens, embedding them in gradient-informed coordinates, and training under partial observation, the Spatial Transformer learns spatial representations that are neurobiologically grounded, robust to noise, and transferable across individuals. This design links the mathematical machinery of attention to established principles of cortical organization, embedding structural–functional hierarchies directly into the model’s representational space. 

% \subsubsection*{Temporal Transformer}
% We treat each time step $t$ as a token by projecting $\mathbf{x}_t\!\in\!\mathbb{R}^{R}$ into $\mathbf{z}_t\!\in\!\mathbb{R}^{D}$, apply a temporal mask and sinusoidal positions~\cite{vaswani2017attention}, and pass through $L$ Transformer blocks. Tokens are decoded to ROI vectors $\hat{\mathbf{x}}_t$; the masked MSE over masked time steps:
% \[
% \mathcal{L}_{\mathrm{temporal}}=\frac{1}{\sum_t (1-M_t)}\sum_t (1-M_t)\|\hat{\mathbf{x}}_t-\mathbf{x}_t\|^2.
% \]
% This captures slow fluctuations and transient events without over-smoothing.

\subsubsection{Temporal Transformer Encoder}

To capture the temporal structure of brain dynamics, BrainSymphony employs a Temporal Transformer that treats each time step as a token and models its dependencies with respect to surrounding context (Fig.~\ref{fig:schematic}a, bottom). This module is complementary to the Spatial Transformer: whereas the spatial encoder learns how regions interact at a given time, the temporal encoder learns how brain-wide activity evolves across time. Training is performed under a masked modeling regime, where a subset of time points is hidden and reconstructed from observed context. This forces the model to capture temporal regularities such as oscillations, cross-regional synchrony, or state transitions.

\textbf{Tokenization and masking.}  
Given an fMRI sample $\mathbf{X} \in \mathbb{R}^{R \times T}$, each time step $t$ is represented by the ROI activation vector $\mathbf{x}_t \in \mathbb{R}^R$. This vector is projected into a $d_f$-dimensional latent space by a learnable projection:
\[
\mathbf{z}_t = \mathbf{W}_{\mathrm{embed}} \mathbf{x}_t, \qquad \mathbf{W}_{\mathrm{embed}} \in \mathbb{R}^{d_f \times R}.
\]
Here, $\mathbf{W}_{\mathrm{embed}}$ is a trainable linear mapping that ensures all time-step tokens are represented in a shared space suitable for attention. To enforce contextual inference, a binary mask $\mathbf{M} \in \{0,1\}^T$ hides a proportion $p$ of time steps. Masked tokens are replaced by a shared learnable embedding $\mathbf{z}_{\mathrm{mask}}$:
\[
\tilde{\mathbf{z}}_t = M_t \cdot \mathbf{z}_t + (1 - M_t) \cdot \mathbf{z}_{\mathrm{mask}}.
\]
This design mimics partial observability in temporal data, teaching the model to reconstruct missing moments from surrounding dynamics.

\textbf{Sinusoidal positional encoding.}  
Temporal order is critical for interpreting brain dynamics. To encode sequence position, we use standard sinusoidal positional embeddings~\cite{vaswani2017attention}:
\[
\mathbf{P}_{t,2i} = \sin\!\left(\frac{t}{10000^{2i/D}}\right), \quad \mathbf{P}_{t,2i+1} = \cos\!\left(\frac{t}{10000^{2i/D}}\right).
\]
The enriched token is then
\[
\mathbf{z}'_t = \tilde{\mathbf{z}}_t + \mathbf{P}_t.
\]
Sinusoidal embeddings provide a smooth and non-learned representation of temporal order, allowing generalization across varying sequence lengths and capturing periodicities naturally. This is particularly useful in fMRI, where dynamics unfold over slow timescales but still exhibit rhythmicity and recurrence.

\textbf{Transformer architecture.}  
The temporally ordered sequence $\{\mathbf{z}'_t\}_{t=1}^T$ is passed through $L$ Transformer blocks, each consisting of multi-head self-attention and feedforward layers with normalization and residual connections:
\begin{align}
\mathbf{h}_t^{(l+1)} &= \mathbf{h}_t^{(l)} + \mathrm{Dropout}\big(\mathrm{MSA}(\mathrm{LN}(\mathbf{h}_t^{(l)}))\big), \\
\mathbf{h}_t^{(l+1)} &= \mathbf{h}_t^{(l+1)} + \mathrm{Dropout}\big(\mathrm{MLP}(\mathrm{LN}(\mathbf{h}_t^{(l+1)}))\big),
\end{align}
with $\mathbf{h}_t^{(0)} = \mathbf{z}'_t$. This structure allows the model to learn both short-range and long-range temporal dependencies by dynamically weighting relationships among time points.

\textbf{Reconstruction objective.}  
Final temporal embeddings $\{\mathbf{h}_t\}_{t=1}^T$ are decoded via a shared MLP $\psi: \mathbb{R}^{d_f} \rightarrow \mathbb{R}^R$ to reconstruct the original ROI activation vectors:
\[
\hat{\mathbf{x}}_t = \psi(\mathbf{h}_t).
\]
The model is optimized with a masked mean squared error loss:
\[
\mathcal{L}_{\mathrm{temporal}} = \frac{1}{\sum_t (1 - M_t)} \sum_{t=1}^T (1 - M_t)\,\|\hat{\mathbf{x}}_t - \mathbf{x}_t\|^2.
\]
This objective encourages the encoder to capture coherent temporal dependencies across ROIs, yielding embeddings that reflect global fluctuations, slow cortical oscillations, task-evoked responses, and dynamic network switching.

By treating time points as tokens, enriching them with sinusoidal order information, and training under masked prediction, the Temporal Transformer acquires representations that are sensitive to the brain’s temporal hierarchy. These embeddings allow BrainSymphony to model both fast and slow dynamics in fMRI, complementing the Spatial Transformer and providing a temporal lens through which brain activity can be understood.

\subsubsection{Context Extractor via 1D Convolution}

Although Transformers excel at modeling long-range dependencies, they can underrepresent short-lived or highly localized dynamics. In fMRI, such dynamics often take the form of transient bursts, fast oscillations, or rapid state transitions that are crucial for understanding cortical processing. To address this, BrainSymphony incorporates a convolutional context extractor that operates independently on each ROI’s time series (Fig.~\ref{fig:schematic}a, right). By applying local temporal filters, this module learns fine-grained features that complement the global dependencies captured by the Transformer encoders. Convolution additionally imposes biologically relevant inductive biases: temporal locality and translation equivariance, reflecting the fact that transient neural events often recur with similar structure at different points in time.

\textbf{Masking strategy.}  
To train this module in a self-supervised manner, we randomly mask a proportion $p$ of entries in the fMRI sequence. Let $\mathbf{M} \in \{0,1\}^{R \times T}$ be a binary mask, where $M_{r,t}=0$ indicates that ROI $r$ at time $t$ is masked. Masked entries are replaced by a shared learnable scalar $\lambda$:
\[
\tilde{x}_{r,t} =
\begin{cases}
x_{r,t}, & \text{if } M_{r,t}=1, \\
\lambda, & \text{if } M_{r,t}=0.
\end{cases}
\]
This preserves input dimensionality while clearly signaling which values are missing, forcing the encoder to infer them from local context.

\textbf{Local temporal encoding.}  
Each masked ROI sequence $\tilde{\mathbf{x}}_r \in \mathbb{R}^T$ is passed through a 1D convolutional encoder:
\[
\mathbf{f}_r = \mathrm{Conv1D}_\theta(\tilde{\mathbf{x}}_r), \qquad \mathbf{f}_r \in \mathbb{R}^{T \times d_c},
\]
where $d_c$ is the number of output channels. The convolution uses a kernel size of 5 with symmetric zero-padding, preserving sequence length while focusing on local neighborhoods of five time points. This design allows the encoder to detect transient fluctuations and oscillatory patterns that span short time windows, complementing the broad-scale dependencies learned by the Temporal Transformer.

\textbf{Decoding and reconstruction.}  
For each ROI $r$ and time point $t$, the contextual feature $\mathbf{f}_{r,t} \in \mathbb{R}^{d_c}$ is decoded by a multilayer perceptron:
\[
\hat{x}_{r,t} = \phi(\mathbf{f}_{r,t}), \qquad \phi:\mathbb{R}^{d_c} \rightarrow \mathbb{R}.
\]
The reconstruction objective focuses only on masked entries.

\textbf{Masked reconstruction loss.}  
Training is performed with a masked mean squared error loss:
\[
\mathcal{L}_{\mathrm{conv}} = \frac{1}{\sum_{r,t}(1-M_{r,t})}\sum_{r=1}^R \sum_{t=1}^T (1-M_{r,t})\big(\hat{x}_{r,t} - x_{r,t}\big)^2.
\]
This objective encourages the model to reconstruct transient neural signals from surrounding local context, ensuring that the learned features reflect temporally coherent structure.

By combining local convolution with masking, this module captures short-lived neural events that global self-attention may overlook. Together with the Spatial and Temporal Transformers, the 1D convolutional extractor ensures that BrainSymphony encodes brain activity across multiple scales—from fast, localized fluctuations to distributed, long-range dependencies—providing a more complete basis for downstream decoding and multimodal fusion.

\subsubsection{Perceiver-based Fusion and Reconstruction}\label{Perceiver}

The three encoding modules described above capture distinct but complementary properties of fMRI signals: spatial dependencies across regions, temporal dynamics across time points, and local fluctuations within short windows. To integrate these representations into a coherent whole, BrainSymphony adopts a fusion mechanism inspired by Perceiver IO~\cite{jaegle2021perceiver}. The key principle is that each ROI–time pair $(r,t)$ carries a unique semantic meaning defined jointly by the above three properties. Instead of processing the full $R \times T$ sequence with direct attention—which would be computationally prohibitive—we leverage a latent cross-attention bottleneck that enables scalable integration while preserving interpretability.

\textbf{Token construction.}  
For every ROI–time pair, we construct a fused token $\mathbf{z}_{r,t}$ by combining three pretrained embeddings. First, ROI embeddings $\mathbf{E}^{\text{ROI}} \in \mathbb{R}^{R \times d_f}$ from the Spatial Transformer encode the anatomical and functional identity of each region. Second, Time embeddings $\mathbf{E}^{\text{Time}} \in \mathbb{R}^{T \times d_f}$ from the Temporal Transformer capture global temporal states. Finally, Context features $\mathbf{F}^{\text{Ctx}} \in \mathbb{R}^{R \times T \times d_c}$ from the 1D convolutional extractor reflect fine-grained local temporal dynamics. 
Each component is projected into a common space and concatenated:
\[
\mathbf{z}_{r,t} = \mathrm{Concat}\!\big(\mathbf{W}_r \mathbf{E}^{\text{ROI}}_r,\; \mathbf{W}_t \mathbf{E}^{\text{Time}}_t,\; \mathbf{W}_c \mathbf{F}^{\text{Ctx}}_{r,t}\big) \in \mathbb{R}^{d_f},
\]
where $\mathbf{W}_r, \mathbf{W}_t, \mathbf{W}_c$ are learnable projections. Intuitively, this token represents “ROI $r$ at time $t$,” enriched with spatial, temporal, and local context.

\textbf{Latent cross-attention.}  
The full set of tokens $\mathbf{Z} \in \mathbb{R}^{RT \times d_f}$ is prohibitively large for direct attention. Instead, we adopt the latent bottleneck of Perceiver IO, introducing $R$ learnable latent vectors $\mathbf{L} \in \mathbb{R}^{R \times d_f}$—one per ROI—that query the fused token space via cross-attention:
\[
\mathbf{L}' = \mathrm{CrossAttn}(\mathbf{L}, \mathbf{Z}).
\]
Each latent $\mathbf{L}'_r$ aggregates all spatiotemporal information relevant to reconstructing ROI $r$’s activity profile, effectively serving as a compact summary of how that region interacts with the rest of the brain over time. A stack of Transformer layers is then applied over these latents to model higher-order interactions among ROIs.

\textbf{Attention Maps for Interpretability.}
The key to the Perceiver's interpretability lies within the \texttt{CrossAttn} operation. This module computes attention scores that quantify the importance of each spatiotemporal input token $\mathbf{z}_{r,t}$ to each latent ROI query $\mathbf{L}_i$. Specifically, the operation first projects the latent queries $\mathbf{L}$ and the fused input tokens $\mathbf{Z}$ into Query ($\mathbf{Q}$), Key ($\mathbf{K}$), and Value ($\mathbf{V}$) spaces:
\begin{align}
    \mathbf{Q} &= \mathbf{L}\mathbf{W}_Q, \qquad \mathbf{Q} \in \mathbb{R}^{R \times d_f} \\
    \mathbf{K} &= \mathbf{Z}\mathbf{W}_K, \qquad \mathbf{K} \in \mathbb{R}^{RT \times d_f} \\
    \mathbf{V} &= \mathbf{Z}\mathbf{W}_V, \qquad \mathbf{V} \in \mathbb{R}^{RT \times d_f}
\end{align}
where $\mathbf{W}_Q, \mathbf{W}_K, \mathbf{W}_V$ are learnable weight matrices.

The \textbf{attention map $\mathbf{A}$} is then computed as the matrix of scaled dot-product scores between queries and keys, normalized by the softmax function:
\begin{equation}
\mathbf{A} = \mathrm{softmax}\left(\frac{\mathbf{Q}\mathbf{K}^\top}{\sqrt{d_f}}\right).
\label{eq:attention_map}
\end{equation}
The shape of this attention map is $\mathbf{A} \in \mathbb{R}^{R \times RT}$. Each entry $A_{i,j}$ represents the weight that the latent query for the $i$-th ROI places on the $j$-th spatiotemporal input token (corresponding to a specific ROI-time pair $(r,t)$).

In other words, the $i$-th row of the attention map, $\mathbf{A}_{i,:} \in \mathbb{R}^{RT}$, is a \textbf{spatiotemporal saliency map for ROI $i$}. It quantifies how much the model ``attends to'' every other ROI $r$ at every time point $t$ when constructing the summary representation for ROI $i$. By reshaping this vector into a matrix of size $R \times T$, we can visualize these attention patterns as dynamic heatmaps. These maps are the key to the model's interpretability, providing a direct, model-derived window into the time-varying functional interactions learned by BrainSymphony. 

To distill these complex spatiotemporal interactions into static, interpretable measures of ROI-level influence, we first marginalize the full attention map $\mathbf{A}$ over the time dimension to create an effective ROI-to-ROI connectivity matrix, $\mathbf{A}' \in \mathbb{R}^{R \times R}$:
\begin{equation}
A'_{i \to j} = \sum_{t=1}^{T} A_{i, (j,t)}.
\label{eq:roi_to_roi_attention}
\end{equation}
Here, $A'_{i \to j}$ represents the total attention that the latent query for ROI $i$ pays to the \textit{entire time series} of ROI $j$. From this, we can define two key metrics:

\textbf{Receptive Attention (ROI-level Inflow).}
For a target ROI $j$, its receptive attention is the total incoming influence it receives from all other ROIs:
\begin{equation}
\mathrm{RA}(j) = \sum_{i \neq j} A'_{i \to j}.
\end{equation}
This metric quantifies how much a given ROI ``listens to'' the rest of the brain when constructing its representation, effectively acting as an information sink.

\textbf{Reconstruction and training.}  
The updated ROI latents $\{\mathbf{L}'_r\}_{r=1}^R$ are decoded back into time series via an MLP:
\[
\hat{\mathbf{x}}_r = \phi(\mathbf{L}'_r), \qquad \phi:\mathbb{R}^{d_f}\rightarrow\mathbb{R}^T,
\]
yielding reconstructed outputs $\hat{\mathbf{X}} \in \mathbb{R}^{R \times T}$. Training minimizes the mean squared error between predicted and original signals:
\[
\mathcal{L}_{\mathrm{perc}} = \frac{1}{RT}\sum_{r=1}^R\sum_{t=1}^T \big(\hat{x}_{r,t} - x_{r,t}\big)^2.
\]
This end-to-end objective ensures that the fused representation retains the full spatiotemporal information necessary to reproduce the observed fMRI signal.

\textbf{Embeddings for downstream tasks.}  
Beyond reconstruction, the ROI-level latents $\{\mathbf{L}'_r\}$ constitute compact, information-rich embeddings that integrate spatial identity, temporal dynamics, and local context. These embeddings can be directly used for downstream tasks such as subject classification, prediction, or condition decoding, providing a concise but comprehensive signature of an individual’s brain dynamics.

This fusion strategy enables BrainSymphony to scale to high-dimensional fMRI data while maintaining interpretability. Each ROI latent can be viewed as an integrated “fingerprint” of how a region behaves across the recording, balancing local detail with global context. By anchoring fusion in cross-attention, the model learns to highlight the most relevant temporal states and local patterns for each region, yielding representations that are simultaneously compact, expressive, and biologically meaningful.

\subsection{Self-Supervised Signed Graph Transformers for Structural Connectomes}
\label{sec:structural-encoder}

While the fMRI branch of BrainSymphony captures dynamic activity patterns, a complete representation of brain function also requires modeling the underlying anatomical scaffold that shapes those dynamics. To this end, we introduce a self-supervised \textit{Signed Graph Transformer} (Fig.~\ref{fig:schematic}b) that learns subject-specific embeddings of the structural connectome. This module departs from conventional graph encoders, which typically assume binary or symmetric adjacency structures, by operating directly on fully weighted connectivity matrices that can be signed and, in principle, asymmetric. Such matrices are typically derived from diffusion MRI tractography and encode the strength of white-matter pathways between regions. Importantly, weights can carry polarity (positive vs.\ negative values), reflecting inhibitory or excitatory influences in effective connectivity estimates, and asymmetry may arise from directional tractography or causal modeling pipelines.

The Signed Graph Transformer learns node embeddings by reconstructing masked edges in the adjacency matrix, thereby forcing the model to internalize both local and global topological features. Its attention mechanism is explicitly sensitive to edge sign and direction, ensuring that positively and negatively weighted links are treated differently during information propagation. 
% In addition, the architecture supports flexible positional encoding schemes that inject graph-theoretic priors—such as node identity, degree, or spectral coordinates—into the latent space. These design choices allow the encoder to construct embeddings that are simultaneously data-driven and biologically interpretable.

By capturing the fine-grained topology of individual structural connectomes, the Signed Graph Transformer produces a structural embedding space that can be seamlessly fused with fMRI-derived representations. This integration provides a multimodal perspective in which temporal brain activity is contextualized by its anatomical substrate, ultimately yielding more robust and neurobiologically grounded subject-level signatures.

\subsubsection{Formulation of structural connectome representation learning}

We formalize the structural branch of BrainSymphony as a graph representation learning task. For each subject $s$, the structural connectome is represented by an adjacency matrix $\mathbf{A}_s \in \mathbb{R}^{R \times R}$, where $R$ is the number of regions of interest (nodes). Each entry $\mathbf{A}_{s,ij}$ encodes the strength of the white-matter connection from node $i$ to node $j$, and may carry both sign and direction. Positive and negative weights allow the model to distinguish between excitatory- and inhibitory-like influences, while asymmetry accommodates directional tractography or effective connectivity estimates.

The goal is to learn node embeddings $\mathbf{H}_s \in \mathbb{R}^{R \times d_g}$ such that the latent representations preserve the structural relationships encoded in $\mathbf{A}_s$. To achieve this without explicit supervision, we adopt a masked edge reconstruction strategy. A subset of edges is randomly hidden during training, and the model must infer their values from the remaining observed structure. Formally, the reconstruction loss penalizes deviations between predicted and true weights only at masked locations.

This setup forces the encoder to capture both local connectivity patterns (e.g., neighborhood structure) and global graph topology (e.g., modularity, hubness, network hierarchy), rather than memorizing the adjacency. In practice, this approach yields embeddings that reflect biologically meaningful structural principles while remaining flexible enough to integrate with fMRI-derived representations in the multimodal fusion stage.

\subsubsection{Model architecture}

To capture subject-specific structural connectomes with biologically informed inductive biases, we design a \textit{Signed Graph Transformer} that directly operates on weighted, signed, and potentially asymmetric adjacency matrices (Fig.~\ref{fig:schematic}b). Unlike conventional graph encoders that often assume undirected or strictly positive edges, our formulation accommodates the full richness of tractography-derived connectivity, where edge weights reflect fiber strength, signs can represent excitatory- versus inhibitory-like influences, and asymmetry can emerge from directional tractography or causal estimates.

Formally, let $\mathbf{A} \in \mathbb{R}^{R \times R}$ denote the structural connectivity matrix for a subject, where $R$ is the number of regions of interest (ROIs). Each node is initially equipped with input features $\mathbf{X} \in \mathbb{R}^{R \times F}$, which may include one-hot identity encodings, degree vectors, or other simple descriptors. These features are projected into a $d_g$-dimensional hidden space:
\[
\mathbf{H}_0 = \mathbf{X} \mathbf{W}_{\mathrm{in}} + \mathbf{P},
\]
where $\mathbf{W}_{\mathrm{in}} \in \mathbb{R}^{F \times d_g}$ is a learnable projection matrix and $\mathbf{P} \in \mathbb{R}^{R \times d_g}$ is a set of learnable node-specific positional embeddings (section~\ref{sec:PE}). This initialization ensures that even in the absence of rich input features, each node retains a unique trainable identity in the latent space.

The node representations are then refined through a stack of $L$ Transformer layers, each conditioned on the adjacency structure:
\[
\mathbf{H}_{\ell+1} = \mathrm{TransformerLayer}(\mathbf{H}_\ell, \mathbf{A}), 
\quad \ell = 0, 1, \ldots, L-1.
\]
Within each layer, multi-head signed attention (section~\ref{sec: signed attention}) governs how information flows between nodes, modulating attention weights according to the magnitude, sign, and direction of structural edges. Iteratively applying these layers yields embeddings $\mathbf{H}_L \in \mathbb{R}^{R \times d_g}$ that integrate both local connectivity patterns and global topological organization of the connectome.

This architecture allows BrainSymphony to move beyond treating the structural connectome as a simple undirected graph. By encoding polarity and directionality in edge-aware attention, the Signed Graph Transformer learns embeddings that are sensitive to the fine-grained organization of anatomical networks, while still capturing higher-order principles such as modularity and hubness. These embeddings form the structural counterpart to the spatiotemporal fMRI representations, enabling meaningful multimodal fusion in later stages.

\subsubsection{Graph-aware multi-head signed attention}
\label{sec: signed attention}

At the core of the Signed Graph Transformer is a modified attention mechanism that explicitly incorporates the polarity and directionality of structural edges (Fig.~\ref{fig:schematic}b, center). Standard Transformer attention distributes weights based only on feature similarity between nodes. In contrast, our graph-aware formulation ensures that attention flows are modulated by the magnitude and sign of anatomical connections, allowing excitatory- and inhibitory-like links, as well as asymmetric pathways, to shape the learned embeddings.

\textbf{Attention projections.}  
For each head $h$, node features $\mathbf{H}_\ell \in \mathbb{R}^{R \times d_h}$ at layer $\ell$ are projected into query, key, and value spaces:
\[
\mathbf{Q}_h = \mathbf{H}_\ell \mathbf{W}_h^Q, \quad
\mathbf{K}_h = \mathbf{H}_\ell \mathbf{W}_h^K, \quad
\mathbf{V}_h = \mathbf{H}_\ell \mathbf{W}_h^V,
\]
where $\mathbf{W}_h^Q, \mathbf{W}_h^K, \mathbf{W}_h^V \in \mathbb{R}^{d \times d_h}$ are learnable projections and $d_h$ is the dimension per head. Raw similarity scores are computed as
\[
\mathbf{S}_h = \frac{1}{\sqrt{d_h}} \mathbf{Q}_h \mathbf{K}_h^\top,
\]
which quantify how strongly node $i$ attends to node $j$ based on their feature representations.

\textbf{Incorporating structural constraints.}  
To respect the connectome’s topology, these similarity scores are modulated by an adjacency-aware weight matrix $\mathbf{W}_{\text{graph}} \in \mathbb{R}^{R \times R}$, derived from the subject’s structural connectivity:
\[
\mathbf{A}_h^{\mathrm{mod}} = \mathrm{sign}(\mathbf{S}_h \odot \mathbf{W}_{\text{graph}})\,\cdot\, 
\mathrm{softmax}\!\left(|\mathbf{S}_h \odot \mathbf{W}_{\text{graph}}|\right).
\]
Here, $\odot$ denotes elementwise multiplication. The magnitude of connectivity modulates attention strength, while the sign ensures that positive and negative links influence embeddings in opposite directions. Asymmetry in $\mathbf{W}_{\text{graph}}$ allows information flow to differ between $i \!\to\! j$ and $j \!\to\! i$, reflecting directional pathways.

\textbf{Message passing.}  
The signed attention weights $\mathbf{A}_h^{\mathrm{mod}}$ are then applied to the values:
\[
\mathbf{Z}_h = \mathbf{A}_h^{\mathrm{mod}} \mathbf{V}_h.
\]
Outputs from all heads are concatenated:
\[
\mathbf{Z} = \mathrm{Concat}(\mathbf{Z}_1, \ldots, \mathbf{Z}_H),
\]
and passed through a residual feed-forward update with normalization:
\[
\mathbf{H}_{\ell+1} = \mathrm{LayerNorm}\!\left(\mathbf{H}_\ell + \mathrm{Dropout}(\mathbf{Z})\right).
\]

This formulation allows the Transformer to treat structural edges not as passive constraints but as active modulators of attention. Positive and negative links drive distinct representational effects, and directionality ensures that influence is not forced to be reciprocal. As a result, the Signed Graph Transformer learns embeddings that respect the nuanced topology of the connectome, capturing both fine-grained local wiring and global organizational principles such as modularity, hub structure, and hierarchy.

\subsubsection{Learnable positional encoding}
\label{sec:PE}

In contrast to temporal sequences, where position corresponds to order, structural graphs require positional information that captures the identity of each region and its role within the network. To provide this inductive bias, we assign every ROI a learnable embedding:
\[
\mathbf{P} = [\mathbf{p}_1; \ldots; \mathbf{p}_R] \in \mathbb{R}^{R \times d_g}, \quad \mathbf{p}_r \sim \mathcal{U}(-\varepsilon, \varepsilon),
\]
where each $\mathbf{p}_r$ is a trainable vector initialized randomly from a uniform distribution $\mathcal{U}(-\varepsilon, \varepsilon)$ and optimized jointly with the model parameters. These embeddings ensure that nodes are distinguishable even when their connectivity profiles are initially similar, and they act as flexible placeholders that can adapt to structural context during training.

This choice contrasts with fixed graph-based encodings (e.g., Laplacian eigenvectors or random walk embeddings), which may impose rigid structural assumptions and suffer from instability across subjects. Learnable node embeddings, provide a universal basis that can be shaped by the optimization objective, enabling the model to capture both invariant properties of specific ROIs (e.g., subcortical hubs) and more subtle topological features (e.g., inter-network bridges). 

These embeddings can be viewed as neurobiologically agnostic “identity tags” that become refined through training to reflect each region’s role in the connectome. Together with the signed attention mechanism, they allow the Signed Graph Transformer to balance anatomical individuality with global organizational principles, yielding node representations that are both flexible and robust across subjects.

\subsubsection{Masked edge reconstruction}

Training the Signed Graph Transformer is formulated as a self-supervised masked edge prediction task, analogous to masked modeling strategies in language and vision~\cite{devlin2019bert,he2022masked}. For each subject, a random subset of edges $\mathcal{E}_{\mathrm{mask}}$ is selected from the structural connectivity matrix $\mathbf{A} \in \mathbb{R}^{R \times R}$. These masked edges are removed from the loss calculation but remain visible to the attention mechanism, ensuring that the model can still leverage the full network context during representation learning.

After $L$ layers of signed graph attention, the final node embeddings $\mathbf{H} \in \mathbb{R}^{R \times d_g}$ are used to reconstruct the missing edges. For a given pair of nodes $i$ and $j$, the predicted weight is obtained via an edge decoder:
\[
\hat{A}_{ij} = f_{\mathrm{edge}}([\mathbf{h}_i;\,\mathbf{h}_j]),
\]
where $[\cdot;\cdot]$ denotes concatenation and $f_{\mathrm{edge}}: \mathbb{R}^{2d_g} \rightarrow \mathbb{R}$ is a multi-layer perceptron. This design allows the decoder to flexibly combine information from both endpoints when estimating the strength and polarity of a connection.

The training objective is a masked mean squared error (MSE) loss:
\[
\mathcal{L}_{\mathrm{graph}} = 
\frac{1}{|\mathcal{E}_{\mathrm{mask}}|} 
\sum_{(i,j)\in \mathcal{E}_{\mathrm{mask}}}
\big(\hat{A}_{ij} - A_{ij}\big)^2.
\]
By focusing only on masked edges, the model is encouraged to infer missing connections from global structural cues rather than memorizing observed entries.

This objective compels the model to learn embeddings that faithfully preserve the topology, polarity, and asymmetry of brain networks. In practice, it forces each ROI’s embedding to encode not only its immediate neighborhood but also higher-order patterns such as modularity, hubness, and inter-network bridges. These rich structural embeddings provide a biologically grounded complement to the spatiotemporal representations from fMRI, ensuring that multimodal fusion in BrainSymphony integrates both dynamic activity and its anatomical substrate.

\subsection{Multimodal fusion of fMRI and structural connectivity representations}
\label{sec:multimodal-fusion}

Functional and structural imaging capture complementary dimensions of brain organization. Functional MRI reflects the dynamic evolution of activity across regions of interest (ROIs), revealing temporal dependencies and state-dependent fluctuations. In contrast, diffusion-derived structural connectivity encodes the white-matter scaffold that constrains these dynamics, shaping both local interactions and large-scale network topology. A multimodal fusion of these signals is therefore essential: it yields representations that are not only sensitive to ongoing neural dynamics but also anchored to the anatomical pathways through which those dynamics unfold (Fig.~\ref{fig:schematic}c, right). This perspective is particularly valuable for capturing individual variability, improving generalization to new populations, and enabling clinically meaningful inference.

To achieve this integration, BrainSymphony introduces a modular fusion framework that combines the latent embeddings learned independently from fMRI (via the spatio-temporal Perceiver module; Fig.~\ref{fig:schematic}a,b) and from structural connectomes (via the Signed Graph Transformer; Fig.~\ref{fig:schematic}c, left). Each ROI is represented by two vectors: a functional embedding $\mathbf{E}_{\mathrm{fMRI}} \in \mathbb{R}^{R \times d_f}$ and a structural embedding $\mathbf{E}_{\mathrm{SC}} \in \mathbb{R}^{R \times d_g}$. Rather than concatenating or averaging these embeddings—which would assume static contributions across regions—we employ an \textit{adaptive gating mechanism} that learns to modulate the relative influence of functional and structural signals for each ROI and each subject (Fig.~\ref{fig:schematic}c, right). This gating layer outputs fused embeddings
\[
\mathbf{E}_{\mathrm{fused}} \in \mathbb{R}^{R \times d},
\]
where $d = d_f = d_g$ is the dimensionality of the shared multimodal space. The fusion weights are optimized end-to-end, allowing the model to discover which modality is most informative for each region under the given task and dataset.
  
This design ensures that the final embeddings simultaneously capture short-term fluctuations in neural activity and the long-range structural constraints that govern them. Regions with strong structural connectivity but weaker dynamic fluctuations (e.g., subcortical hubs) can be grounded by anatomical priors, while regions with highly state-dependent dynamics (e.g., prefrontal cortex) can emphasize functional signals. By flexibly adjusting to task and subject demands, BrainSymphony’s adaptive fusion produces node-level representations that are both functionally expressive and anatomically grounded, enabling superior performance on downstream tasks and offering new opportunities for neuroscientific interpretability.

\subsubsection{Formulation of multimodal fusion}

Let $\mathbf{E}_{\mathrm{fMRI}} \in \mathbb{R}^{B \times R \times d_f}$ denote the fMRI-derived embeddings for a batch of $B$ subjects across $R$ regions of interest (ROIs), where $d_f$ is the embedding dimensionality. These embeddings are produced by the Perceiver-based fusion of spatial, temporal, and contextual features described in Section~\ref{sec:spatiotemporal} (Fig.~\ref{fig:schematic}a,b). Similarly, let $\mathbf{E}_{\mathrm{SC}} \in \mathbb{R}^{B \times R \times d_g}$ denote the node embeddings obtained from the Signed Graph Transformer operating on structural connectivity data (Section~\ref{sec:structural-encoder}, Fig.~\ref{fig:schematic}c, left). 

The objective is to integrate these two modality-specific representations into a joint multimodal embedding:
\[
\mathbf{E}_{\mathrm{fused}} \in \mathbb{R}^{B \times R \times d},
\]
where $d$ is the dimensionality of the shared latent space. The fused representation is then passed to downstream modules for tasks such as sex classification, age prediction, or unsupervised network identification.

A naive strategy would be to concatenate $\mathbf{E}_{\mathrm{fMRI}}$ and $\mathbf{E}_{\mathrm{SC}}$ or average them. However, such approaches assume that both modalities contribute equally across all brain regions and subjects. In reality, the balance between structure and function is highly heterogeneous: in some regions, stable anatomical constraints dominate, while in others, dynamic functional fluctuations carry more predictive information. To address this, BrainSymphony employs an \textit{adaptive gating mechanism} (Fig.~\ref{fig:schematic}c, right). This mechanism learns a soft, region-specific weighting of the two modalities, enabling the model to dynamically adjust the contribution of fMRI versus structural embeddings for each ROI and each subject. 
  
The adaptive gate can be viewed as a biologically inspired filter that determines, for every region, whether its representation should lean more heavily on dynamic activity or on anatomical scaffolding. This flexibility allows BrainSymphony to align embeddings from fundamentally different representational spaces and to generate a coherent, unified latent representation. In doing so, it captures both subject-specific variability and universal organizational principles—an essential property for building robust, generalizable multimodal brain models.

\subsubsection{Adaptive gating fusion.}  
To integrate the functional and structural embeddings into a unified representation, we introduce an adaptive gating mechanism that learns, for each region and subject, how much weight to assign to fMRI- versus SC-derived features. Formally, the fused embedding at each ROI is defined as a convex combination of the two modality-specific embeddings:
\begin{equation}
\mathbf{E}_{\mathrm{fused}} = 
\sigma\!\left(\mathbf{W}_g [\mathbf{E}_{\mathrm{fMRI}} \,\|\, \mathbf{E}_{\mathrm{SC}}]\right) 
\odot \mathbf{E}_{\mathrm{fMRI}} \;+\; 
\Big(1 - \sigma\!\left(\mathbf{W}_g [\mathbf{E}_{\mathrm{fMRI}} \,\|\, \mathbf{E}_{\mathrm{SC}}]\right)\Big) 
\odot \mathbf{E}_{\mathrm{SC}},
\end{equation}
where $\|$ denotes concatenation along the feature dimension, $\sigma(\cdot)$ is the element-wise sigmoid function, and $\mathbf{W}_g \in \mathbb{R}^{2d \times d}$ is a learnable projection matrix. The gate values are bounded between 0 and 1, ensuring that the fused embedding remains a convex combination of the two modalities.

This formulation is motivated by the observation that the influence of structure and function is neither static nor uniform across the brain. In primary sensory cortices, structural wiring tends to exert a stronger constraint on activity patterns, whereas in higher-order association regions, flexible and state-dependent dynamics may dominate. The balance between modalities can also vary across individuals, contexts, and clinical conditions. By learning region-specific and subject-specific gates, the model is able to capture this heterogeneity and adaptively emphasize the modality most informative for a given setting.  

In contrast to naive fusion strategies such as simple concatenation or static averaging, adaptive gating introduces both flexibility and interpretability. The gating mechanism allows the contribution of each modality to vary across space and subjects, effectively down-weighting noisy or less reliable signals when appropriate. At the same time, the gate values themselves provide a window into neurobiological organization: they can reveal which regions rely more heavily on structural scaffolds and which depend more on dynamic functional fluctuations. This dual role, as both an integrative operator and an interpretable output, makes adaptive gating a powerful strategy for multimodal fusion in neuroimaging.

\subsection{Baselines}

To rigorously evaluate BrainSymphony, we benchmarked it against a diverse set of established and state-of-the-art models spanning convolutional, graph-based, and foundation-style architectures. These baselines were chosen to reflect the major methodological families currently employed in neuroimaging representation learning, thereby ensuring that our comparisons capture both domain-specific and general-purpose advances.

\begin{itemize}
\item \textit{BrainNetCNN}~\cite{kawahara2017brainnetcnn} is a pioneering convolutional architecture specifically designed for brain connectivity matrices. It introduced edge-to-edge and edge-to-node convolutions to capture localized patterns in structural and functional connectomes.  
 \item \textit{BrainGNN}~\cite{li2021braingnn} extends this line of work by applying graph neural networks (GNNs) with ROI-aware pooling, enabling hierarchical feature extraction that preserves anatomical interpretability.  
 \item \textit{Brain Network Transformer (BNT)}~\cite{kan2022brain} adapts the Transformer framework to brain graphs, explicitly modeling long-range dependencies across nodes, and thus represents the Transformer family tailored for connectomic data.  
 \item \textit{Vanilla Transformer (VT)} serves as a strong deep learning baseline, applying the standard Transformer architecture directly to brain data without domain-specific modifications.  
 \item \textit{BrainLM}~\cite{caro2023brainlm} exemplifies large-scale foundation models for fMRI, employing a masked autoencoding objective to learn generalizable time-series embeddings.  
Finally, 
 \item \textit{Brain-JEPA}~\cite{dong2024brain} represents the latest generation of foundation models in neuroimaging, leveraging a Joint Embedding Predictive Architecture (JEPA) to predict masked data in latent space, thereby capturing rich spatiotemporal representations.
\end{itemize}

Together, these baselines span the evolution of the field: from early CNN-based methods optimized for connectome data, to graph-based approaches incorporating structural priors, to Transformer models that leverage long-range dependencies, and finally to large-scale self-supervised foundation models. Benchmarking BrainSymphony against this spectrum of approaches ensures that its advantages cannot be attributed to narrow methodological choices but instead reflect genuine progress in multimodal and parameter-efficient brain modeling.

% \subsection*{Evaluation metrics and statistics}
% For classification we report Accuracy and F1; for regression we report MSE and Pearson correlation ($\rho$). We compute subject-level metrics on held-out test sets, with validation for early stopping. Where applicable, we provide 95\% bootstrap confidence intervals (10{,}000 resamples) and effect sizes, and we control multiple comparisons (FDR) in ROI-wise analyses. Statistical procedures, resampling details, and additional robustness checks (e.g., ablations of fusion gate; removal of a component; unimodal-only) are in Supplementary~\S S8.

% \subsection{Implementation details}
% \textcolor{red}{Models were implemented in PyTorch with mixed precision. Unless otherwise stated, $D{=}128$, $d_\ell{=}128$, $L{=}6$ for Transformers, and $H{=}8$ attention heads. AdamW with cosine decay and warmup was used for pretraining and fine-tuning. Hardware, compute budgets, training time, and code release information are in Supplementary~\S S9.}

\section*{Code Availability}
The Python implementation of BrainSymphony—supporting pretraining on new datasets—together with pretrained model checkpoints (frozen weights) is available at the following GitHub repository: \href{https://github.com/Moein-Khajehnejad/BrainSymphony}{GitHub
 repository}.

\section*{Author contributions}
M.K.\ and A.R.\ conceived the study and designed the experiments. M.K.\ performed the experiments, analysed the data, and contributed materials/analysis tools. F.H.\ performed the experiments and analysed the data. M.K.\, F.H.\, A.R.\, and D.S.\ contributed to the writing of the manuscript. All authors reviewed, revised, and approved the final manuscript.

% \bibliography{sn-bibliography}% common bib file

\clearpage
% \section*{Supplementary Figures and Tables}
% \begin{itemize}
%     \item Fig.~S1: Distribution of streamline lengths and weights for SC matrices.
%     \item Fig.~S2: Gradient alignment across subjects (before/after Procrustes).
%     \item Fig.~S3: Ablation results of fMRI encoder streams.
%     \item Fig.~S4: Comparison of Graph positional encodings (learnable vs RWPE).
%     \item Fig.~S5: Fusion strategy comparison (concatenation, averaging, gating).
%     \item Fig.~S6: Attention entropy distributions across components.
%     \item Fig.~S7: Reconstruction fidelity (R$^2$ distributions across ROIs).
%     \item Table~S1: Asymmetry robustness of Signed Graph Transformer.
%     \item Table~S2: Hyperparameters summary.
%     \item Table~S3: Parameter counts of baselines.
% \end{itemize}

\newcounter{extendeddata}
\newcommand{\beginsupplement}{
    % To number supplemental material with 'E'
    \renewcommand{\thepage}{S\arabic{page}} 
    \renewcommand{\thesection}{S\arabic{section}}
    \renewcommand{\thetable}{S\arabic{table}}  
    \renewcommand{\thefigure}{S\arabic{figure}}
    %\renewcommand{\figurename}{Extended Data Fig.} %text used to open the fig caption
    %\renewcommand{\bibnumfmt}[1]{[S#1]}
    %\renewcommand{\citenumfont}[1]{S#1}
    % reset counters
    \setcounter{page}{1}
    \setcounter{section}{0}
    \setcounter{table}{0}
    \setcounter{figure}{0}
    \setcounter{equation}{0}
    \newcounter{SIfig}
    \renewcommand{\theSIfig}{\arabic{SIfig}}
    }

% \begin{appendices}
\beginsupplement
\makeatletter
\renewcommand{\thesection}{S\arabic{section}}
\renewcommand{\theHsection}{S\arabic{section}}
\renewcommand{\thetable}{S\arabic{table}}
\renewcommand{\theHtable}{S\arabic{table}}
\renewcommand{\thefigure}{S\arabic{figure}}
\renewcommand{\theHfigure}{S\arabic{figure}}
\renewcommand{\theequation}{S\arabic{equation}}
\renewcommand{\theHequation}{S\arabic{equation}}
\makeatother

\section*{Supplemental Materials}
\label{supp_mat}
This Supplementary Information provides additional details on datasets, preprocessing, model architecture, training, and evaluation that support the results in the main text. We also report extended figures and implementation details for reproducibility. 

\section{Dataset preprocessing}\label{sup_dataset}

\subsection{Functional MRI}\label{sup_fMRI}
For HCP-YA and HCP-Aging, we used minimally preprocessed resting-state fMRI provided by the HCP consortium~\cite{elam2021human}. Preprocessing included gradient unwarping, motion correction, B0 distortion correction, and ICA-FIX denoising. For PsiConnect, preprocessing followed Novelli et al.~\cite{novelli2025psiconnect}. 

All datasets were parcellated into 450 ROIs (Schaefer-400 cortical parcels~\cite{schaefer2018local} and Tian-Scale III subcortical parcels~\cite{tian2020topographic}). Signals were detrended, band-pass filtered (0.008–0.09 Hz), and nuisance-regressed (motion parameters, WM/CSF signals). Time points with framewise displacement >0.5 mm were scrubbed. Each ROI’s time series was robust scaled (subtract median, divide by IQR) across participants, following BrainLM~\cite{caro2023brainlm}.

To ensure reproducibility and compatibility with the pretrained \textit{BrainSymphony} framework, users should adhere to the same data preparation and parcellation conventions employed during model development. The model was trained using functional MRI (fMRI) time series that were robustly scaled per region of interest (ROI) across all participants according to the transformation:
\[
x' = \frac{x - \text{median}(x)}{\text{IQR}(x)},
\]
where $\text{IQR}(x)$ denotes the interquartile range. This normalization step attempts to mitigate scanner- and site-specific intensity variations while preserving the relative temporal structure of each signal.

Input data must follow the same parcellation schema, comprising a total of \textbf{450 ROIs} organized as:  
(1--50) \textit{Tian-Scale III subcortical regions},  
(51--250) \textit{Schaefer-400 cortical regions (left hemisphere)}, and  
(251--450) \textit{Schaefer-400 cortical regions (right hemisphere)}.  
The fMRI inputs are expected in tensor form $(N, R, T)$, where $N$ denotes the number of subjects, $R=450$ the number of ROIs, and $T$ the number of time points per subject (default = 200). For each subject, the structural connectivity matrix derived from diffusion MRI (dMRI) tractography should be aligned to the same ROI order.

\subsection{Structural connectivity}\label{sup_SC}
Diffusion MRI preprocessing used MRtrix3: denoising, motion/eddy-current correction, bias-field correction, and spherical deconvolution tractography. SC matrices were constructed as streamline counts normalized by ROI volume, log10-transformed, and aggregated into the 450-ROI parcellation. Resulting matrices are \emph{symmetric, weighted, and signed} (negative weights arise from z-scoring). Tracts shorter than 20 mm and longer than 250 mm were discarded. 

BrainSymphony was designed with a modular architecture, allowing flexible use of its multimodal encoders. Depending on data availability, users may provide either the functional component (fMRI encoder), the structural component (Signed Graph Transformer), or both jointly for multimodal fusion. The model’s pretrained encoders can be loaded directly, while the downstream prediction head (e.g., regression, classification, or reconstruction) can be fine-tuned to the specific task at hand. Fine-tuning typically involves updating only the task-specific layers, keeping the pretrained encoder weights frozen to preserve the learned neurobiological priors.

\section{Training protocol and hyperparameters} \label{hyperparams}
Table~\ref{tab:arch_summary} summarizes the key architectural choices and training hyperparameters used in this work. We provide details for each major component of the model, including spatial and temporal Transformer encoders, the Conv1D signal encoder, the Perceiver module, the signed graph Transformer, the adaptive fusion layer, and the final MLP head, as well as training configuration. Unless otherwise specified, the same settings were applied consistently across all datasets and tasks.

\begin{table}[t]
\centering
\caption{Summary of architectural choices and hyperparameters for the proposed model.}
\label{tab:arch_summary}
\begin{tabularx}{\textwidth}{l l Y}
\toprule
\textbf{Module} & \textbf{Field} & \textbf{Setting / Value} \\
\midrule

\multirow{3}{*}{\textbf{Data}}
& ROIs ($R$) & 450 \\
& Time steps ($T$) & 200  \\

\midrule
\multirow{5}{*}{\textbf{Spatial Transformer}}
& Dim ($d_f$) & 128\\
& Layers / Heads & 6 / 4 \\
& Positional Encoding & Learnable + Gradient \\
& MLP Ratio & 4.0 \\
& Dropout & 0.2 \\
& Masking ratio &  70\%  \\

\midrule
\multirow{5}{*}{\textbf{Temporal Transformer}}
& Dim ($d_f$) & 128 \\
& Layers / Heads & 6 / 4\\
& Positional Encoding & Sinusoidal \\
& MLP Ratio & 4.0 \\
& Dropout & 0.2 \\
& Masking ratio &  70\%  \\

\midrule
\multirow{4}{*}{\textbf{Conv1D Encoder}}
& Filters ($d_c$) & 64 \\
& Kernel size & 5 \\
& Padding & 2 \\
& Dropout & 0.2 \\
& Masking ratio &  50\%  \\

\midrule
\multirow{5}{*}{\textbf{Perceiver}}
& Latents ($L$) & 450  \\
& Dim ($d_f$) & 128\\
& Blocks & 1 \\
& Heads & 2 \\
& Dropout & 0.2 \\

\midrule
\multirow{5}{*}{\textbf{Signed Graph Transformer}}
& Dim ($d_g$) & 128 \\
& Layers / Heads & 8 / 4  \\
& Positional Encoding & Learnable \\
& Graph type & weighted, signed  \\
& Dropout & 0.2 \\

\midrule
\multirow{3}{*}{\textbf{Fusion}}
& Fusion Type & Adaptive Gating \\
& Fusion Dim ($d$) & 128 \\

\midrule
\multirow{6}{*}{\textbf{Training}}
& Optimizer & AdamW, lr $=1\times 10^{-4}$ \\
& Scheduler & Cosine decay\\
& Warmup &  10\% of total training steps \\
& Batch size / Epochs & 16 / 1500 \\
& Weight Decay & 0.01 \\
& Seed Splits &  70/15/15 \\

\bottomrule
\end{tabularx}
\end{table}

% \subsection{Pretraining}
% All encoders were pretrained jointly with loss
% \[
% \mathcal{L} = \lambda_s \mathcal{L}_{\text{spatial}} + \lambda_t \mathcal{L}_{\text{temporal}} + \lambda_c \mathcal{L}_{\text{conv}} + \lambda_p \mathcal{L}_{\text{perc}} + \lambda_g \mathcal{L}_{\text{graph}}.
% \]
% Weights were $\lambda_s=\lambda_t=\lambda_c=\lambda_p=1.0$, $\lambda_g=0.5$.  
% Masked proportion $p=0.3$ for spatial and temporal, $p=0.2$ for CNN.  

% \subsection{Optimization}
% Optimizer: AdamW ($\beta_1=0.9,\beta_2=0.999$, weight decay $=10^{-2}$).  
% Learning rate: 1e-4, cosine decay with 10k warmup steps.  
% Batch size: 8 subjects.  
% Training epochs: 100 (pretraining), 50 (fine-tuning).  

% \subsection{Architecture}
% Embedding dimension $D=128$, latent dimension $d_\ell=128$, hidden MLP dimension $512$, heads $H=8$, layers $L=6$. Context CNN: 1D conv with kernel size 5, stride 1, padding 2, output channels $d_c=64$.  

% Table~S2 summarizes all hyperparameters.

% \section{Fine-tuning and evaluation}

% \subsection{Linear probing vs fine-tuning}
% For HCP-Aging, we split subjects into train/val/test (6:2:2). For linear probing, encoders were frozen and only a linear head trained. For fine-tuning, all parameters updated. Early stopping based on validation loss.  

\section{Component-specific attention patterns reveal complementary mechanisms}\label{att_maps}

\begin{figure}[h!]
    \centering
    \includegraphics[width=0.9\linewidth]{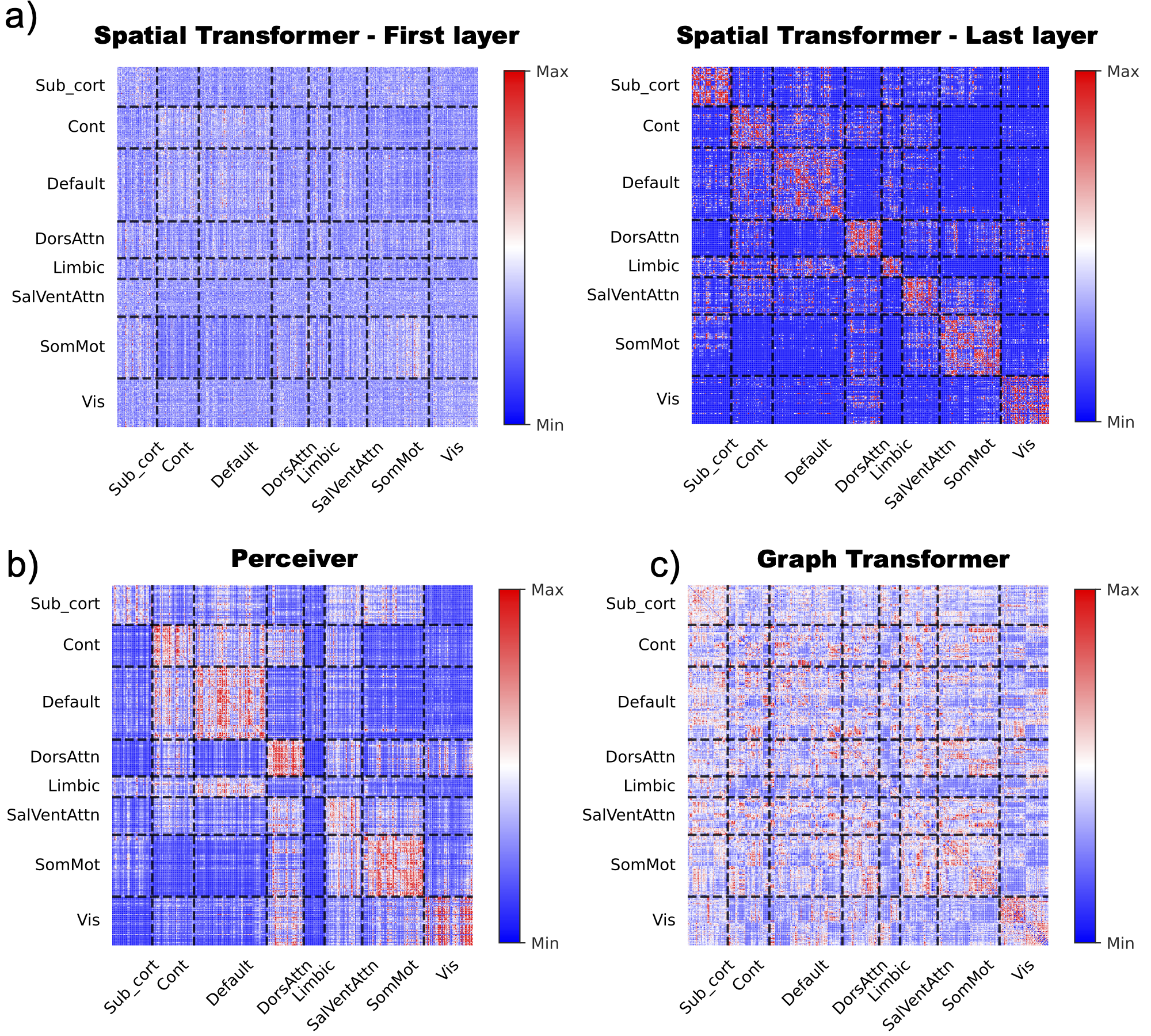}
    \caption{\textbf{Component-wise attention patterns capture complementary dependencies.}
    Each heatmap shows average pairwise attention between 450 ROIs, grouped by eight functional networks (dashed boundaries).
    \textbf{a)} Spatial Transformer: early layers show broad interactions; deeper layers emphasize intra-network blocks with selective inter-network links.
    \textbf{b)} Perceiver: integration of high-dimensional spatiotemporal inputs into ROI-specific embeddings highlights regions central for global integration.
    \textbf{c)} Graph Transformer: attention patterns reflect graph-aware relations consistent with known macroscale organization. Color scales are normalized within component.}
    \label{fig:S1}
\end{figure}

Beyond benchmark performance, it is essential to understand how different architectural components of BrainSymphony contribute to its representations. To this end, we examined the average pairwise attention weights from the Spatial Transformer, Perceiver, and Graph Transformer modules, sorted by eight canonical functional networks (\nameref{ex_data} Fig.~\ref{fig:S1}). These maps provide a mechanistic window into how the model organizes dependencies across the brain.

The Spatial Transformer exhibited a clear progression across layers. Early layers showed diffuse, relatively homogeneous attention across regions (\nameref{ex_data} Fig.~\ref{fig:S1}a, left), consistent with an initialization stage where the model considers broad, unspecialized spatial dependencies. By the final layer, attention patterns sharpened into stronger diagonal blocks, reflecting increased intra-network cohesion, while simultaneously highlighting selective cross-network links (\nameref{ex_data} Fig.~\ref{fig:S1}a, right). This evolution indicates that the Spatial Transformer gradually refines diffuse information into more specialized and neurobiologically meaningful network structure.

The Perceiver module, tasked with compressing high-dimensional spatiotemporal input into ROI-specific embeddings, emphasized regions central for integrating across systems (\nameref{ex_data} Fig.~\ref{fig:S1}b). Its attention maps illustrate how the Perceiver balances preservation of network identity with the need for global communication, allowing compact but information-rich embeddings. This mechanism helps explain why BrainSymphony can operate efficiently on massive spatiotemporal data without losing key integrative features.

The Graph Transformer, by design, incorporates structural connectivity and thus produces attention patterns that closely align with known macroscale brain architecture (\nameref{ex_data} Fig.~\ref{fig:S1}c). Its maps highlighted intra-network associations but also emphasized hierarchical inter-network dependencies, reflecting the topology of the connectome. This suggests that the Graph Transformer not only encodes the presence of structural links but also exploits their organization to inform functional embedding.

Together, these component-specific analyses demonstrate that BrainSymphony achieves interpretability by distributing representational work across modules with complementary roles: the Spatial Transformer evolves from global to refined spatial relations, the Perceiver integrates high-dimensional signals into compact global embeddings, and the Graph Transformer leverages anatomical priors to anchor learned relations in structural connectivity. This multi-view complementarity underpins both the model’s predictive power and its alignment with known principles of brain organization, showing how architectural design choices translate into neuroscientific interpretability.

\section{Multiclass decoding of attention maps across cognitive states}
\label{attention_decoding}

To assess whether the learned attention maps encode meaningful and discriminative information about ongoing cognitive states, we designed a multiclass classification framework spanning four experimental conditions: \emph{rest}, \emph{meditation}, \emph{music}, and \emph{movie}.

\paragraph{Methodological Approach.}  
For each subject, we first extracted the full ROI-by-ROI attention matrices ($450 \times 450$). To avoid trivial self-connections, diagonal entries were excluded. Each matrix was then vectorized into a feature vector, yielding a subject-level representation of attentional interactions.  

To reduce dimensionality and control for feature scaling, we implemented a linear pipeline consisting of:
\begin{enumerate}
    \item \textbf{Z-scoring (Standardization):} Each feature was standardized across subjects.
    \item \textbf{Principal Component Analysis (PCA):} Components explaining $95\%$ of the variance were retained, ensuring compact but information-rich representations.
    \item \textbf{Logistic Regression Classifier:} A linear classifier was trained on the reduced feature set.
\end{enumerate}

Performance was quantified using a leave-one-subject-out cross-validation (LOSO-CV) scheme, ensuring that classification was evaluated strictly on unseen participants. Metrics included overall accuracy (ACC), balanced accuracy (BACC; mean per-class recall), and macro-averaged F1 score. Confusion matrices were used to visualize condition-specific decoding patterns.

\begin{figure}[h!]
    \centering
    \includegraphics[width=0.9\linewidth]{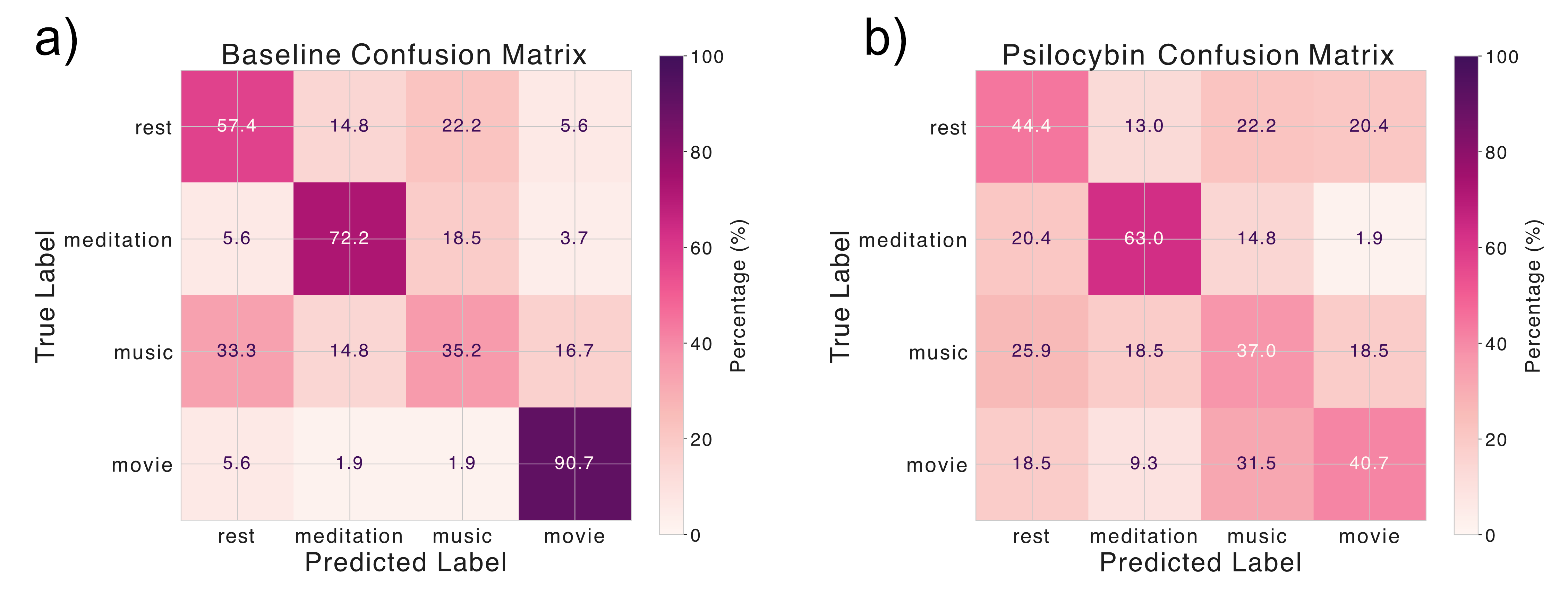}
    \caption{\textbf{Imaging condition decoding from BrainSymphony attention maps.}
\textbf{a)} Baseline (no psilocybin) confusion matrix showing accurate classification of Rest, Meditation, Music, and Movie conditions based on ROI–ROI attention features, with the highest precision for Movie and Meditation.
\textbf{b)} Following psilocybin administration, condition-specific decoding accuracy decreased, reflecting a flattening of categorical structure and increased overlap between conditions. This is consistent with the globally integrated attention patterns observed under psilocybin (see Fig.~\ref{fig:3}).}
    \label{fig:S4}
\end{figure}

\paragraph{Results — Baseline}  
The classifier achieved \textbf{ACC = 0.639}, \textbf{BACC = 0.639}, and \textbf{F1-macro = 0.628} across 216 held-out subject samples.
\begin{itemize}
    \item The confusion matrix (Fig.~\ref{fig:S4}a) revealed strong discriminability for \emph{movie} and \emph{rest}, both exceeding $\sim$75\% correct classification.
    \item By contrast, \emph{music} was often confused with \emph{rest} and \emph{meditation}, suggesting that attentional organization during music shares overlapping features with internally focused states.
    \item The relatively high performance indicates that attention maps in the baseline condition reliably capture state-dependent distinctions, especially between conditions with distinct sensory or cognitive demands (external movie viewing vs.\ internally oriented rest).
\end{itemize}
These findings suggest that the structure of attention maps is not arbitrary, but encodes separable cognitive fingerprints at the network level.

\paragraph{Results — Psilocybin}  
In contrast, classification under the psilocybin condition yielded lower performance (\textbf{ACC = 0.463}, \textbf{BACC = 0.463}, \textbf{F1-macro = 0.463}).
\begin{itemize}
    \item The confusion matrix (Fig.~\ref{fig:S4}b) showed more evenly distributed misclassifications, with no condition standing out as consistently separable.
    \item In particular, \emph{music} and \emph{movie} were more frequently confused, and \emph{rest} was only modestly discriminable from the other states.
\end{itemize}
This reduction in separability suggests that the psilocybin administration is associated with a flattening of attentional distinctions, potentially reflecting either (i) greater heterogeneity across individuals, or (ii) a global shift in network organization that reduces the specificity of condition-linked patterns.

\paragraph{Interpretation.}  
Together, these results support the notion that attention maps carry informative structure about cognitive states, but that their discriminative power depends on context:
\begin{itemize}
    \item In the baseline setting, attention maps reveal clear state-specific signatures, particularly for externally engaging vs.\ internally oriented states.
    \item After the psilocybin administration, the reduced classification accuracy suggests either a blurring of these signatures or increased variability across subjects.
\end{itemize}
Importantly, this approach provides a quantitative validation that attention maps are not random or artifactual, but encode condition-relevant information at the network level.

\section{Comparison with functional connectivity analyses}\label{fc_comp}

\begin{figure}[h!]
    \centering
    \includegraphics[width=0.9\linewidth]{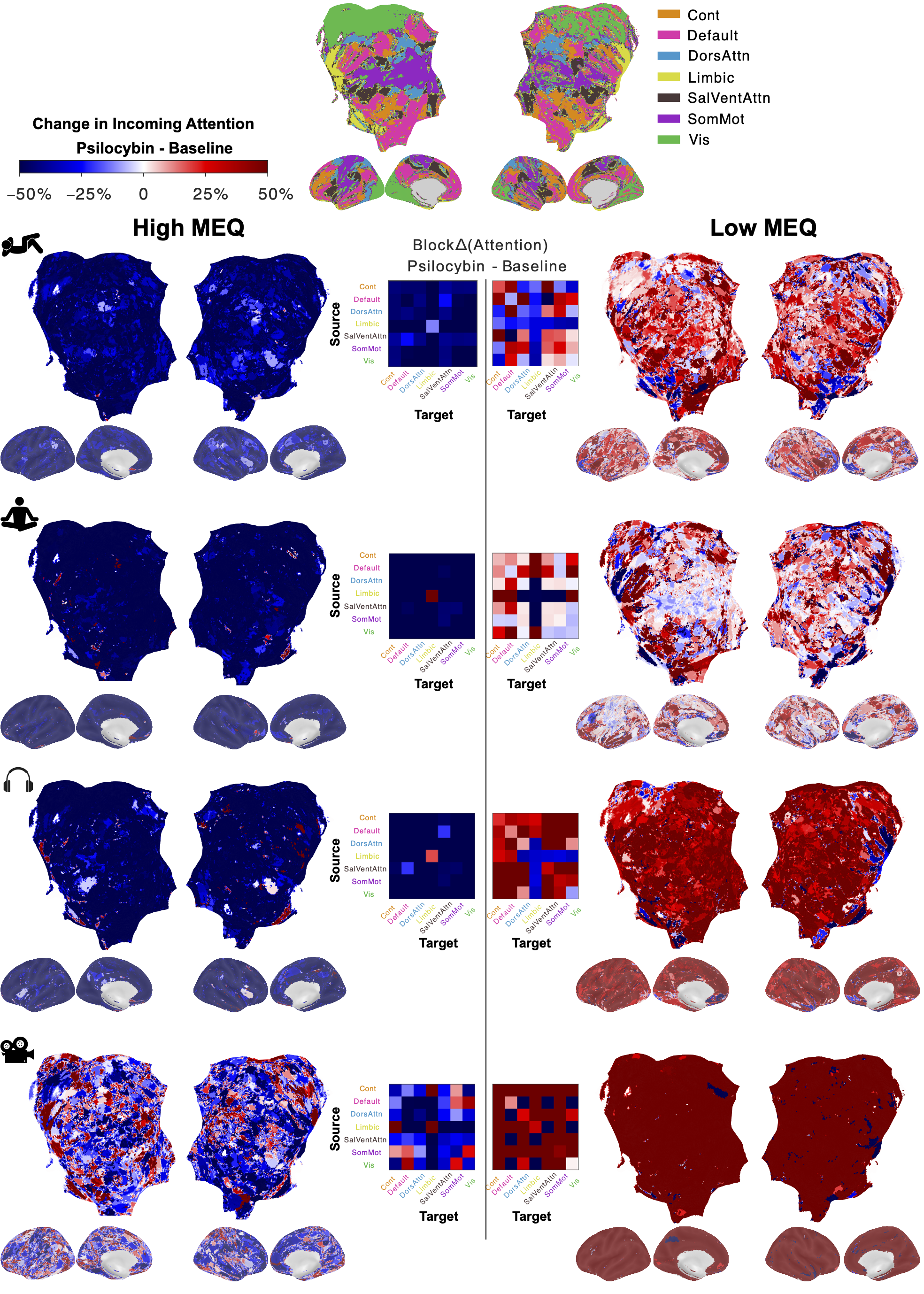}
    \caption{\textbf{Psilocybin-related changes in functional connectivity across MEQ subgroups and conditions.}
Top row: reference flatmaps showing the seven canonical functional networks. Remaining rows: difference maps (Psilocybin – Baseline) of Pearson correlation-based functional connectivity for subjects in the top (High MEQ, left) and bottom (Low MEQ, right) deciles of MEQ30 scores. Each row corresponds to one condition from the PsiConnect dataset (rest, meditation, music, movie). Heatmaps (center) display block-averaged changes between networks. While high-MEQ individuals showed widespread decreases in connectivity across conditions, low-MEQ individuals exhibited global increases. Only in meditation, music, and movie conditions did high-MEQ subjects show localized inter-network increases involving limbic regions, partially echoing the attention-based findings but in a far more limited fashion.}
    \label{fig:S7}
\end{figure}

For completeness, we performed an analogous analysis to that of Section \ref{psiconnect}
using conventional functional connectivity (Pearson correlation) rather than attention-based receptive inflow (\nameref{ex_data} Fig.~\ref{fig:S7}). Here, high-MEQ participants showed widespread decreases in connectivity across nearly all conditions under psilocybin, whereas low-MEQ participants exhibited global increases. Unlike the attention results, these effects were largely undifferentiated and did not reveal region-specific or mechanistic patterns that could explain the intensity of subjective experience. One partial convergence with the attention findings was observed in meditation, music, and movie conditions, where high-MEQ subjects showed modest inter-network increases involving the limbic system. However, these effects were much weaker and less nuanced than those captured by attention maps, underscoring the added interpretive value of BrainSymphony’s mechanistic embeddings compared to traditional connectivity metrics.

% \section{Implementation and compute}

% Models implemented in PyTorch 2.0. Training used 4 NVIDIA A100 GPUs (40 GB). Pretraining on HCP-YA+40\% HCP-Aging took 72 GPU-hours. Fine-tuning per downstream task <5 GPU-hours.  

\section*{Extended Data} \label{ex_data}
% -----------------------------
% Figure S1 caption 
% -----------------------------

\begin{figure}[h!]
    \centering
    \includegraphics[width=1\linewidth]{figures/Figure_S2.png}
    \caption{\textbf{Distribution and example reconstructions of fMRI time series across conditions.} For each condition (rows: Rest, Meditation, Music, Movie) and drug state (columns: Psilocybin vs. Baseline), histograms show the distribution of ROI-wise reconstruction metrics: coefficient of determination ($R^2$, purple), mean absolute error (MAE, grey), and Pearson correlation $\rho$ (teal). Dashed red lines indicate mean values. Rightmost panels display representative ROI time series comparing original BOLD (black) and model reconstruction (red), illustrating the model’s performance across temporal scales.
    }
    \label{fig:S2}
\end{figure}

\begin{figure}[h!]
    \centering
    \includegraphics[width=1\linewidth]{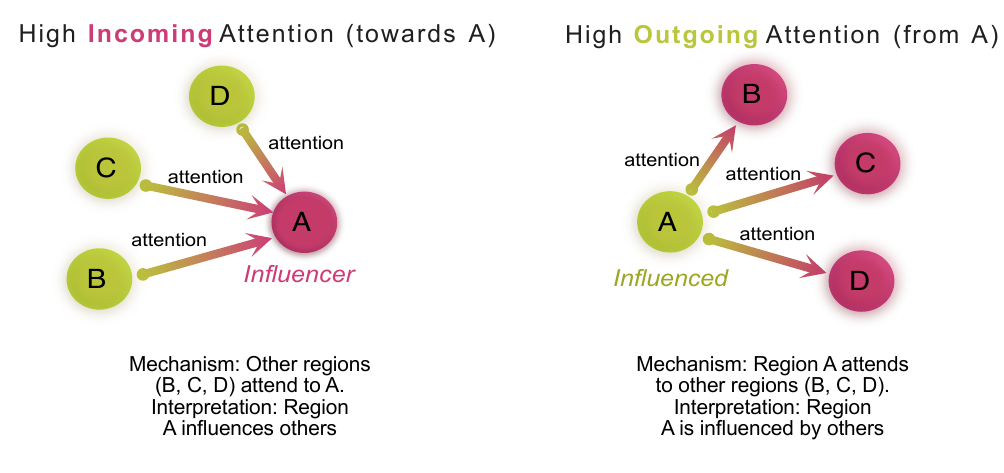}
\caption{Schematic clarifying the relationship between attention direction and influence in BrainSymphony.
\textbf{(Left)} High incoming attention (to A) signifies that A is an "influencer" region.
\textbf{(Right)} High outgoing attention (from A) signifies that A is an "influenced" region.}
\label{fig:S3}
\end{figure}

\begin{figure}[h!]
    \centering
    \includegraphics[width=0.9\linewidth]{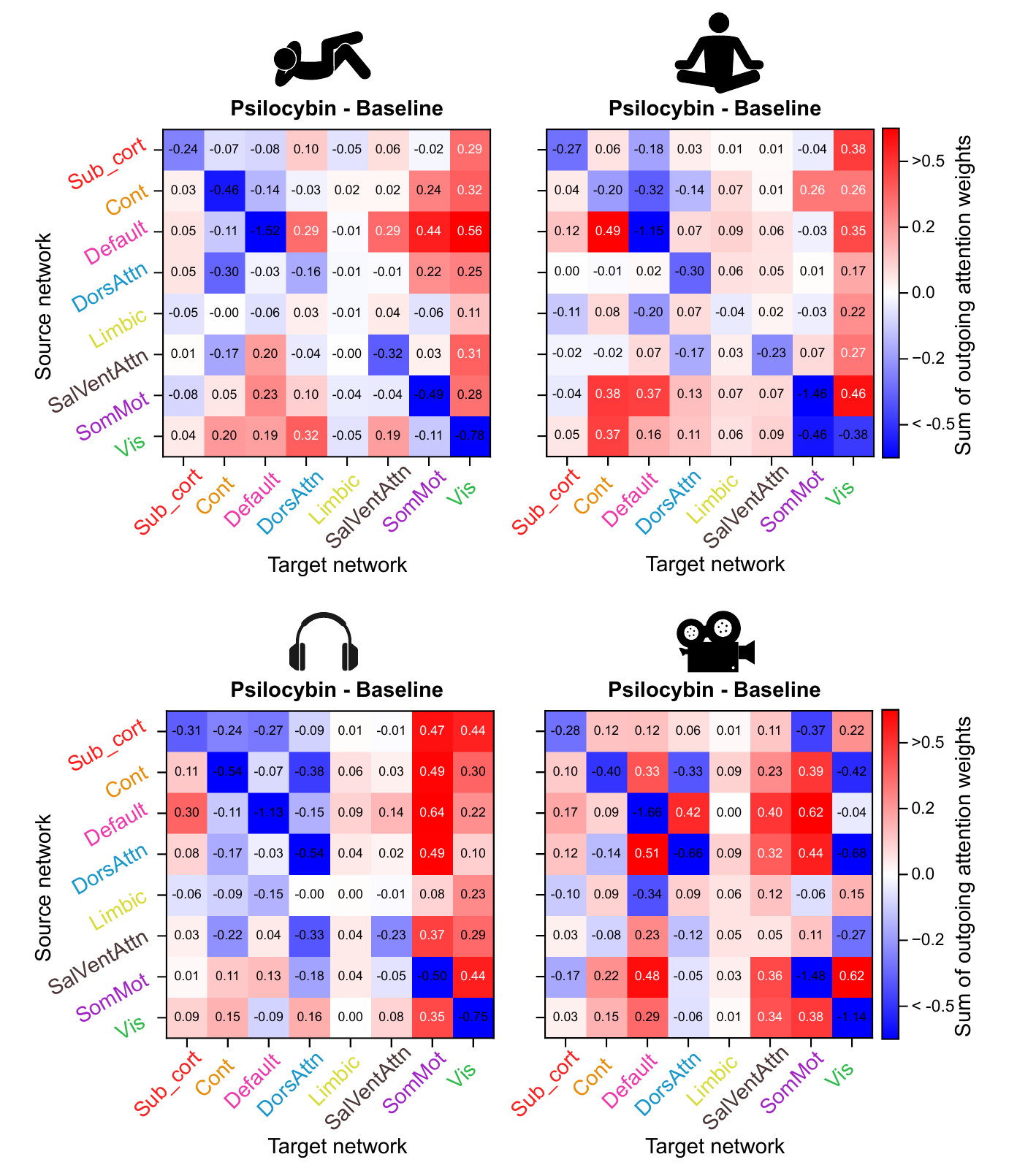}
    \caption{\textbf{Changes in network-level outgoing attention weights across imaging conditions under psilocybin versus baseline.}
    Each matrix depicts the difference in summed outgoing attention weights between psilocybin and baseline states (Psilocybin - Baseline), averaged across all subjects and aggregated within ROIs of each canonical functional network. The four panels correspond to distinct imaging conditions — Rest, Meditation, Music listening, and Movie viewing. Source networks (rows) indicate the origin of directed attention, while target networks (columns) indicate the recipient networks (influencers). Warm colors (red) represent increased outgoing attention under psilocybin relative to baseline, whereas cool colors (blue) indicate decreased outgoing attention.}
    \label{fig:S5}
\end{figure}

\begin{figure}[h!]
    \centering
    \includegraphics[width=0.9\linewidth]{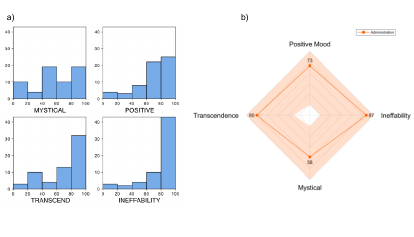}
    \caption{\textbf{MEQ-30 score distributions and factor profile for the Psiconnect sample.}
\textbf{a)} Histograms showing the distribution of scores across the four validated MEQ-30 factors (Mystical, Positive Mood, Transcendence, and Ineffability).
\textbf{b)} Radar plot illustrating the average factor scores, highlighting high ratings for ineffability and transcendence consistent with strong mystical-type experiences.}
    \label{fig:S6}
\end{figure}

% \subsection{Reconstruction fidelity}
% We computed R$^2$ between reconstructed vs original time series across ROIs. Mean R$^2$=0.81±0.07 across subjects (Fig.~S7). Structural reconstructions preserved within-network modularity (Q=0.37 vs 0.39 in ground truth).  

\clearpage




% \end{appendices}


\end{document}